\documentclass[11pt]{article}
\usepackage[margin=1in]{geometry}
\usepackage[centertags]{amsmath}
\usepackage{amsfonts,amsthm}
\usepackage{rotating}
\usepackage{multirow}
\usepackage{subfigure}
\usepackage{kbordermatrix}
\usepackage[all]{xy}
\usepackage{graphicx}

\usepackage[small, bf, margin=25pt, tableposition=top]{caption}

\newcommand\Tgap{\rule{0pt}{2.6ex}}
\newcommand\Bgap{\rule[-1.2ex]{0pt}{0pt}}
\newcommand{\E}{\mathbf{E}}
\newcommand{\R}{\mathcal{R}}
\newcommand{\Py}{\mathcal{P}}
\newcommand{\CS}{\mathcal{S}}
\newcommand{\T}{\mathcal{T}}

\newtheorem{thm}{Theorem}
\newtheorem{cor}{Corollary}
\newtheorem{lem}{Lemma}

\theoremstyle{definition}

\theoremstyle{remark}
\newtheorem{rem}{Remark}

\begin{document}
\title{On the Imitation Strategy for Games on Graphs}

\date{}
\author{Colin Cooper \thanks{Department of Computer Science, King's College, University of London, London WC2R 2LS, UK}
\and
Martin Dyer \thanks{School of Computing, University of Leeds, Leeds LS2 9AS, UK}
\and
Velumailum Mohanaraj \footnotemark[2]}

\maketitle

\abstract{In  evolutionary game theory, repeated two-player games are
used to study strategy evolution in a population under natural
selection. As the evolution greatly depends on the interaction structure,
there has been growing interests in studying the games on graphs. In
this setting, players occupy the vertices of a graph and play the game only
with their immediate neighbours. Various evolutionary dynamics
have been studied in this  setting for different games. Due to the
complexity of the analysis, however, most of the work in this area is experimental.
This paper aims to contribute to a more complete understanding, by providing
rigorous analysis. We study the imitation dynamics on two classes of graph: cycles and complete graphs.
We focus on three well known social dilemmas,  namely the
Prisoner's Dilemma, the Stag Hunt and the Snowdrift Game. We also consider, for completeness,
the so-called Harmony Game. Our analysis shows that, on the cycle,
all four games converge fast, either to total cooperation or total defection.
On the complete graph, all but the Snowdrift game converge fast, either to cooperation
or defection. The Snowdrift game reaches a metastable state fast, where
cooperators and defectors coexist. It will converge
to cooperation or defection only after spending time in this state which is exponential
in the size, $n$, of the graph. In exceptional cases, it will remain in this state indefinitely. Our theoretical results are
supported by experimental investigations.}


\allowdisplaybreaks
\section{Introduction}

Game theory is used as a mathematical tool to analyse strategic and
economic decision-making by rational agents~\cite{Ohtsuki2006}.
Evolutionary game theory, on the other hand, studies the evolution of
strategies  in these situations by natural selection. Thus it models the adaptive behaviour of agents who meet repeatedly.
Different strategic situations can be modelled by simple games, for example repeated $ 2 \times 2$ games
with appropriate payoff matrices (see, for example, \cite{Kilgour1988,Snidal1991}).
This is the setting we consider here.

In this context, there are two players, each having two pure strategies (\emph{Cooperate} and \emph{Defect}). Here we study \emph{symmetric} games.
A game is said to be symmetric
if it puts each player in the same strategic situation. Each player's
payoff depends on the strategies chosen by the player and the opponent. The payoff obtained when both players cooperate is denoted by
$R$ (the \emph{reward for mutual cooperation}). The payoff gained when both defect is denoted
by  $P$ (the \emph{punishment for mutual defection}). Finally, $T$ (the \emph{temptation to defect})  is
earned by the defector and $S$  (the \emph{sucker's payoff}) is earned by the cooperator when
one player defects and the other cooperates. Hence, the payoff matrix is:
\begin{displaymath}
\kbordermatrix{
   & Cooperate & Defect \\
Cooperate & R & S \\
Defect & T & P}\ .
\end{displaymath}

Assuming strict ordinality of the payoffs (i.e.\ all four payoffs are distinct)  gives rise to 12 different strategic
games that are symmetric~\cite{Kilgour1988}.  In order to select the most relevant games, some more realistic assumptions are made about the payoffs. First,
it is assumed that unilateral defection is preferred to unilateral
cooperation (i.e.\ $T > S$), thus there is an incentive for non-cooperative
behaviour~\cite{Snidal1991}. Furthermore, for a dilemma to arise, the following
conditions need to be satisfied~\cite{Michael2002}: (1) mutual cooperation is preferred over mutual defection (i.e.\ $R > P$), (2)
mutual cooperation is preferred to unilateral cooperation (i.e.\ $R>S$), and (3) unilateral defection is preferred over mutual cooperation (i.e.\ $T>R$) or
mutual defection is preferred over unilateral cooperation (i.e.\ $P>S$). All these assumptions yield four strategic
situations. They are : $T>R>S>P$ which represents the Hawk-Dove or \emph{Snowdrift} game (SG); $R>T>P>S$ or $R>P>T>S$
which represents the Assurance or \emph{Stag-Hunt} game (SH)~\cite{Michael2002}; and $T>R>P>S$ which represents the \emph{Prisoner's Dilemma} game (PD).
In addition to these well known social dilemmas, another game is widely studied which models the situation
where the interests of the players match. This is represented by  the regime defined by the rankings $R>S>T>P$~\cite{Licht1999}
and $R>T>S>P$~\cite{Snidal1991}. This game is called \emph{Harmony} game (HG). In  this game, cooperating strictly dominates
defecting so that the only possible equilibrium
of the game is for both players to cooperate. As this equilibrium is Pareto-optimal, there is no dilemma in this game.

\begin{figure}[ht]
\centering
\subfigure[Categorisation based on the game dynamics and the games.] {
\label{fig:gamedomainsNotLabelled}
  \includegraphics[width=35mm, angle=-90]{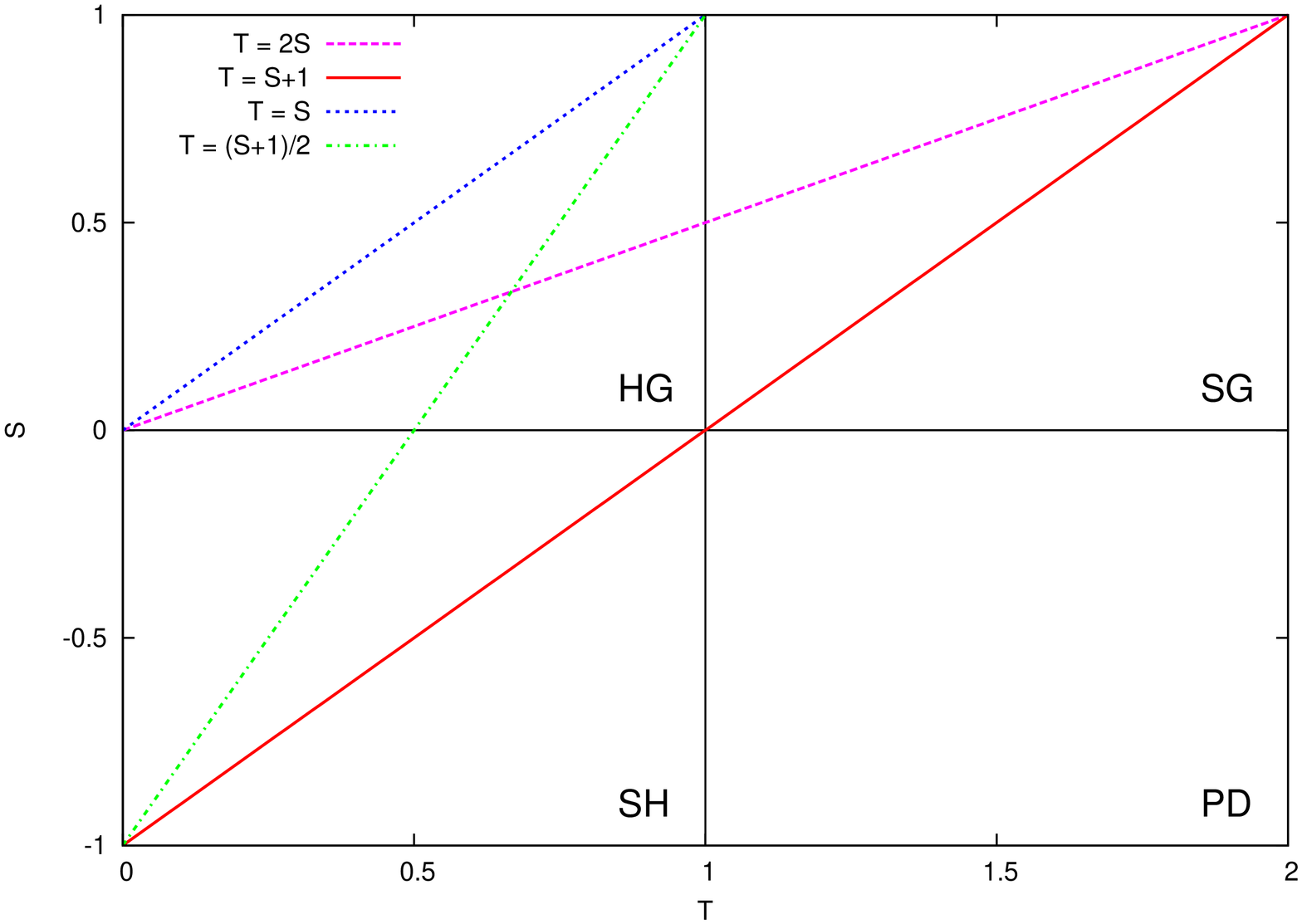}}%
\hspace{1cm}
\subfigure[Categorisation based on the game dynamics.] {
\label{fig:gamedomainsLabelled}
  \includegraphics[width=35mm, angle=-90]{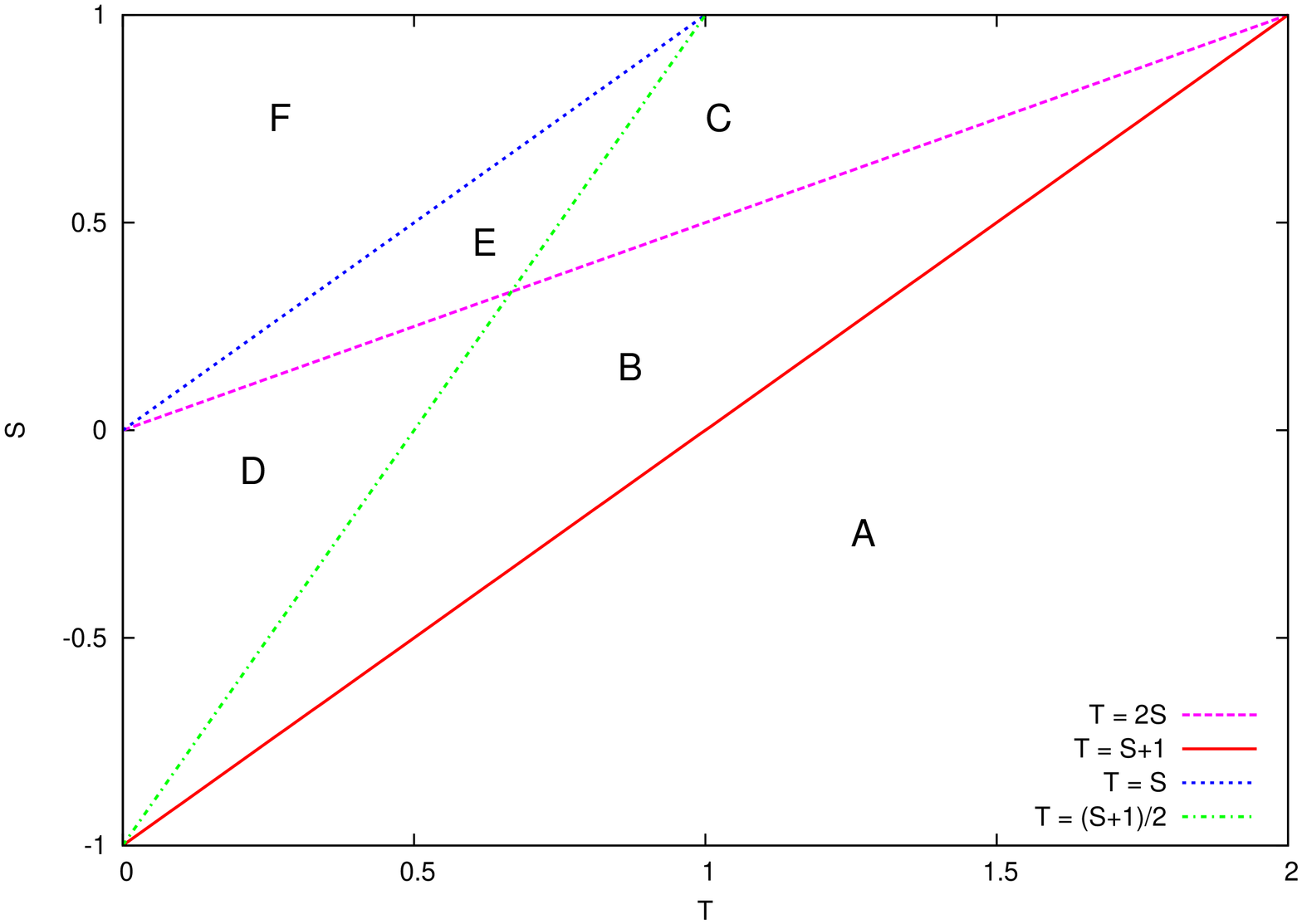}}%
\caption[Game domains.]
{Game domains.}
\label{fig:gamedomains} 
\end{figure}

We use the normalisation of \cite{Santos2006a} to simplify the analysis. That is, we normalise the payoff for mutual
cooperation ($R$) to 1 and the payoff for mutual defection ($P$) to 0.
Then, as done in \cite{Santos2006a}, if we restrict  $ 0 \le T \le 2$ and $-1 \le S \le 1$, the behaviour of all four games
can be captured, with each game corresponding to a quadrant in the $ST$ plane as shown in Figure~\ref{fig:gamedomainsNotLabelled}.
Note that the quadrant for SH includes only the  version defined by $R>T>P>S$ which is the standard version studied
in the literature (see, for example, \cite{Michael2002,Roca2009a,santos2006}). Thus the other version
defined by $R>P>T>S$ is omitted in this study.

Understanding the evolution of cooperation among selfish agents is
clearly an important and challenging task. Much effort has been put into
achieving this using the PD game as a model~\cite{martin1992}.
When the population is mixed, where each player is equally likely to meet any
other, natural selection favours defection
over cooperation~\cite{Nowak2006}. Hence, Nowak and May \cite{martin1992}
studied the impact of arranging PD players in a two-dimensional array
and concluded that cooperators and defectors can coexist indefinitely.
Since then, considerable attention has been given to studying evolutionary
game dynamics in spatial settings. In these settings, players are arranged
as the vertices of a network and can play the game only with their
immediate neighbours. The impact of network on the
emergence of cooperation has also been emphasised in  \cite{kittock93emergent,pachco2005,santos2006}.
The way in which cooperation evolves in spatial settings is called \emph{network reciprocity} \cite{Nowak2006}, where cooperators
survive by forming a cluster and helping each other within the cluster
so that the defectors at the border cannot fare any better.
In this paper we consider two extreme cases of the spatial setting.
First, we consider the cycle graph in which the impact of the topology
on the evolution is strongest~\cite{Ohtsuki2006}. Second, we consider the complete graph which
models a mixed population. Several previous studies have focused on these
types of graphs (see, for example, \cite{dyer00convergence,kittock93emergent,Ohtsuki2006,Santos2006a}).

In evolutionary game theory, the payoffs are regarded as the Darwinian fitness~\cite{Ohtsuki2006}.
During the evolution, strategies earning higher payoffs become more
common in the population. Inheritance and imitation are two mechanisms
by which successful strategies may spread. Between the two,
imitation gives the more practical dynamics~\cite[p.~86]{hofbaur83}.
Many versions of imitation have been studied in this context.
Nowak and May \cite{martin1992} studied the imitation rule known as \emph{unconditional-imitation} or
\emph{imitate-the-best}. Here, each player imitates the neighbour
earning the highest payoff among the immediate neighbours and himself, in each round of the game.
Furthermore, one of the  three update rules studied in \cite{Ohtsuki2006} is the asynchronous version
of the \emph{proportional-imitation}~\cite[p.~87]{hofbaur83} rule. Under this rule,
in each round of the game, a random individual is given a chance to update
his strategy. The individual then chooses a neighbour uniformly at random and imitates the
neighbour with some probability proportional to the payoff difference. (A simpler version
of this update rule is called \emph{imitate the better}~\cite[p.~87]{hofbaur83}, in which an
updating individual always imitates the randomly selected neighbour, but only if the neighbour's
payoff is higher.)
The synchronous version
of this has been studied in \cite{Roca2009b,Santos2006a}.  In this rule, each individual
updates his strategy at the end of each round of the game in this fashion.
(\cite{Roca2009b} refers to this as the \emph{replicator rule}.) Finally,
a stochastic combination of both versions
of the proportional-imitation rule has been studied in \cite{Roca2009a}.
Among the variations of imitation update rule, proportional-imitation rules perform
optimally, both from the individual's perspective and from the perspective of
 the population as a whole~\cite{Schlag1998}.
In this paper, we will study the synchronous version of the proportional-imitation rule.
In the rest of this paper, we frequently refer to this rule just as the imitation update rule.

Most of the previous work on imitation rules has been empirical.
For instance,  the imitate-the-best rule on a two dimensional grid was studied in~\cite{martin1992}
using simulations; the synchronous proportional-imitation rule on different types of graphs was explored in~\cite{Roca2009b,Santos2006a}
using simulations; and both of these imitation rules were investigated in  \cite{Roca2009a}  using simulations. The reason for
the lack of rigorous analysis is that a vast number of possible patterns of strategies can be generated~\cite{Nowak2006}.
The empirical results give insights into the evolution, but some
of the results cannot properly be understood without theoretical
underpinning. On the theoretical side,  the asynchronous version of the
proportional-imitation-rule on the cycle
was analysed in \cite{Ohtsuki2006} using \emph{fixation probabilities}. The fixation probability is the
probability that a population adopting the same strategy is overrun by a single
individual adopting a mutant strategy. Although the results presented in \cite{Ohtsuki2006} are
interesting, the analysis based on fixation probabilities has two weaknesses.
First, it does not show what happens when mutants invade in pairs, triples, etc. Second,
it does not reveal any information about the rate of convergence to cooperation.

Here we will study the synchronous proportional-imitation rule on
cycles and complete graphs \emph{rigorously}. Similar rigorous results were
given for the Pavlov or Highest Cumulative Reward rule in the Iterated PD game (see \cite{dyer00convergence,mossel2007} for details).
Here, we make no assumptions about the initial configuration. We then calculate the time it takes for a
steady state to be reached in these settings. By doing so, we provide
rigorous support for the experimental results observed by \cite{Santos2006a} in
complete graphs. In addition, we do a similar study for the cycle. Interestingly, this simple
type of graph gives evidence that there are
graphs on which cooperators and defectors cannot coexist for any of
the four games, except for some specific payoffs values (i.e.\ $T=S+1$ and $T=(S+1)/2$). Furthermore we
provide supporting simulation results for both types of graph.

The outline of this paper is as follows: Some preliminaries are described in Section~\ref{s:preliminaries}.
Section~\ref{sec:cycle} investigates the dynamics of the imitation on the
cycle, while Section~\ref{sec:completeGraph} investigates the same on the complete
graph. Empirical results appear in Section~\ref{sec:simulation}.
The impact of our results is discussed in Section~\ref{sec:discussion}.
Finally, concluding remarks are presented in Section~\ref{s:conclusion}.

\section{Preliminaries} \label{s:preliminaries}

Let $G=(V,E)$ be a connected graph with vertices
$V = \{v_i: i=0,1,\ldots,n-1\}.$
Each player at the vertex $i $ $(0  \leq   i  \leq n-1)$
has a state $\CS_i \in \{0, 1\}$, where $0$ represents \emph{defection} and $1$ represents
\emph{cooperation}. In addition, $\star$ is used as the \emph{don't care} symbol for the states when the particular value
of the state does not matter in the discussion (e.g.\  $0\star1000$).

We will now define formally the \emph{proportional-imitation} rule \cite{hofbaur83} with synchronous update.
According to this rule, in each generation, each vertex $v_i$ ($ 0
\le i < n$) plays the game with all its $\mathcal{N}_i$ neighbours, and stores the accrued payoff for that
generation as
$P_i$. At the end of the generation, all vertices simultaneously
update their strategies as follows: each vertex $v_i$ chooses one of
its neighbours u.a.r.\footnote{uniformly at random} (say $v_{j}$) and copies the strategy of $v_j$ with probability
\begin{equation*}
 p_i = \max \left\{0, \dfrac{P_{j} - P_{i}}{m\alpha} \right\},
\end{equation*}
\noindent where $\alpha =  \max\{T, 1\}- \min\{S,0\}$, and
$m = \max\{\mathcal{N}_i, \, \mathcal{N}_j\}$.  The denominator $m\alpha$  is a scaling factor
which ensures that $p_i \in [0,1]$. Here, $p_i$ is called the \emph{switching probability} of
$v_i$. Note that if $P_{i} \ge P_{j}$, $v_i$ keeps the same strategy, i.e.\ the switching
probability  $p_i = 0$. Clearly, the value of $m$ is 2 for cycles and $n-1$ for complete graphs.

We are interested to find the \emph{absorption time} which is defined as the time required
for the system to reach a steady state. The absorption time is determined in terms
of the number of generations it takes for absorption as a function of the number
of players, $n$. All-cooperate (where all players cooperate) and
all-defect (where all players defect) are clearly two steady states under the imitation
update rule.
And there may exist other steady state configurations, where cooperators and defectors can coexist. Once the system reaches any of these states, it will remain there forever.

\section{Imitation on the cycle}
\label{sec:cycle}

Suppose that a minimum of three agents occupy the vertices of a cycle graph $G=(V,E)$, where
\begin{equation*}
V = \{v_i: i=0,1,\ldots,n-1\}\ \ \textrm{and}\ \ E=\{\{v_i,v_{i+1}\}:i=0,1,\ldots,n-1\}\ .
\end{equation*}
Here and throughout this paper, addition and subtraction on the vertex
subscripts is performed modulo $n$. We now introduce some terminology. Let $\mathcal{S} \in \{0, 1\}^V$ be given. A \emph{cooperator-run}
(resp. defector-run) in $\mathcal{S}$ is an interval $[i, j]$  where $0 \leq i, j < n$, such that
$S_k = 1$ (resp. $0$) for $i\leq k \leq j$ and $\CS_{i - 1} = 0$ (resp. 1), $\CS_{j + 1} = 0$
(resp. 1). We use \emph{c-run} to denote the cooperator-run and \emph{d-run}
to denote the defector-run. It is possible to have $j \le i$, since we are working modulo $n$.
Clearly all runs are disjoint.

Suppose that $\R_\textrm{d} = [i, j]$ is a d-run where the subscript d stands for defectors. The
length of the run $\R_\textrm{d}$, denoted by $\ell(\R_\textrm{d})$, is the number of defectors in the
run, i.e.\ $\ell(R_\textrm{d})=j-i+1\mod n$. We will refer to a d-run of length $\ell$ as an
$\ell_\textrm{d}$-run. A $1_\textrm{d}$-run will also be called a \emph{singleton} defector and a
$2_\textrm{d}$-run will also be called a \emph{pair} of defectors. There are two \emph{outer rim}
edges associated with $\R_\textrm{d}$, namely $\{i - 1, i\}$ and $\{j, j+1\}$. We use similar
notations for c-runs, but with subscript c, which stands for cooperators. For example,
$\ell_\textrm{c}$-run represents a c-run of
length $\ell$. A run, c-run or d-run, is said to \emph{grow} if its length increases. A
run whose length cannot be reduced to zero through any combination of updates is
called a \emph{barrier}. A run is said to be \emph{deleted} if all its vertices change
from defector to cooperator or vice versa.

The main results for the game on cycle are listed below. The proofs
are presented in the subsequent sections.

\bigskip

\noindent \textbf{Initial configuration:~} We assume that the initial configuration
for the game is random: each vertex on the cycle is independently assigned as a cooperator
with constant probability $p_\textrm{c}$ and as a defector with probability $1-p_\textrm{c}$
at the beginning of the game. Then, the following theorems give the absorption time for the
$n$-vertex cycle.

\begin{thm}
\label{thm:fastconvA}
If the payoffs are such that $S+1 < T$, then the game converges to the
all-defect state in time $O(n)$  with high probability.
\end{thm}


\begin{thm}
\label{thm:fastconvBandCfinal}
If the payoffs satisfy
\begin{description}
    \item [Case I:] $\tfrac{S+1}{2} < T < S+1$ and $T>2S,$ or
    \item [Case II:] $\tfrac{S+1}{2} < T < 2S$,
\end{description}
then the game converges to the all-cooperate state in time $O(n)$  with
high probability.
\end{thm}


\begin{thm}
\label{thm:fastconvDEandFfinal}
If the payoffs satisfy
\begin{description}
    \item [Case I:] $0 < T < \tfrac{S+1}{2}$  and $T>2S$, or $S < T < 2S$  and $T < \tfrac{S+1}{2}$, or
    \item [Case II:] $T < S$,
\end{description}
then the game converges to the all-cooperate state in time $O(\log n)$  with high probability.
\end{thm}

\medskip

\begin{rem}
In this paper, an event $Y_n$ which depends on the size of the graph $n$ is said to happen
\emph{ with high probability}, or in short \emph{w.h.p.}, if $\Pr(Y_n) \to 1$ as $n \to \infty $.
\end{rem}

\subsection{Analysis}
\label{sec:analysis}

Suppose, without loss of generality, that the random neighbour chosen by $v_i$
for imitation is $v_{i+1}$. Then,
the switching probabilities of $v_i$ for different values of $\CS_{i}$ and $\CS_{i+1}$
are given in Figure~\ref{fig:switchingprob}.
Clearly, we can ignore the two cases where the updating vertex and
the randomly chosen neighbour already follow the same strategy, hence imitation
has no effect. Ignoring these cases and
expanding the formulae in Figure~\ref{fig:switchingprob} with the possible values of $\CS_{i-1}$ and
$\CS_{i+2}$, we obtain the results in Figure~\ref{fig:switchingprob2}. This figure
also shows the variable names we use to denote these different probabilities. Now, it is
obvious that the switching probability of $v_i$, when it randomly chooses $v_{i+1}$ for
imitation, depends on the states $\CS_{i-1},\CS_i,\CS_{i+1}$, and $\CS_{i+2}$.
For notational convenience
we write these states as  $\CS_{i-1}[\CS_i]\CS_{i+1}\CS_{i+2}$ (e.g.\  $0[1]11$), enclosing
the state of $v_i$ in square brackets.

\begin{figure}
\centering
\begin{tabular}{|c|c|l|}
  \hline
  $\CS_{i}$ \Tgap& $\CS_{i+1}$ & $p_i= \tfrac{P_{i+1} - P_{i} \Tgap}{2\alpha}$ \\
   [2ex] \hline
   0& 0 & $\tfrac{(\CS_{i+2} -  \CS_{i-1})T  \Tgap}{2\alpha}$ \\
   [2ex] \hline
    0 & 1 &$\tfrac{(2S-T) + \CS_{i+2}(1-S) - \CS_{i-1 }T  \Tgap}{2\alpha}$ \\
   [2ex] \hline
    1 & 0 & $\tfrac{(T-2S) + \CS_{i+2}T - \CS_{i-1}(1-S)  \Tgap}{2\alpha \Bgap}$ \\
   [2ex] \hline
    1 & 1 & $\tfrac{(\CS_{i+2} -  \CS_{i-1})(1-S) \Tgap}{2\alpha \Bgap}$ \\
  \hline
\end{tabular}
\caption[Switching probabilities for  $v_i$.]{Switching probabilities for  $v_i$, provided $v_{i+1}$ has been chosen by $v_i$ for imitation.}
\label{fig:switchingprob} 
\end{figure}

Note that the prerequisite for a player to switch strategy through imitation is to have at least one neighbour
with a different strategy. Hence, the strategy changes on the cycle can happen only at the
vertices of the outer rim edges of a c-run or a d-run, since all other vertices incident on a
run have both their neighbours employing the same strategy as theirs. So, it is sufficient to focus our analysis
on the borders of runs.

\begin{figure}
\centering
\begin{tabular}{|c|c|c|c|c|c|}
  \hline
 \multirow{2}{*}{$\CS_{i}$}  \Tgap & \multirow{2}{*}{$\CS_{i+1}$}& \multicolumn{4}{|c|}{$\CS_{i-1}\CS_{i+2}$} \\
\cline{3-6}
  &  & 00 & 01 & 10 & 11\\
\hline
    0 & 1 & $a = \tfrac{2S-T\Tgap}{2\alpha }$ & $b = \tfrac{S+1-T}{2\alpha}$  & $c = \tfrac{2(S-T)}{2\alpha}$  & $d =\tfrac{S+1-2T}{2\alpha}$  \\
   [1ex] \hline
        1 & 0 & $\bar{a} = \tfrac{T-2S\Tgap }{2\alpha \Bgap}$ & $\bar{c} = \tfrac{2(T-S)}{2\alpha}$  & $ \bar{b} = \tfrac{T-S-1}{2\alpha}$  & $ \bar{d} = \tfrac{2T-S-1}{2\alpha}$  \\
[1ex]
  \hline
\end{tabular}
\caption[Switching probabilities for  $v_i$.]{Switching probabilities for  $v_i$ for different possible neighbour-states.}
\label{fig:switchingprob2} 
\end{figure}

For the analysis, we first need to know whether the values for the $p_i$'s are
zero or not. As mentioned earlier, each game corresponds to a quadrant in the $ST$ plane as shown in Figure~\ref{fig:gamedomainsNotLabelled}.
Yet, we will not use this game-based categorisation in our analysis. This is because,
although these games have different
($\alpha, S, T$) values, they share the same behaviour in some cases, as a
result of having the same sign  for $P_i - P_j$\,:

\begin{itemize}
\item $T > \tfrac{S+1}{2}$ for PD and SG
\item $T > 2S$ for PD and SH
\item $T > S$ for PD, SH and SG
\end{itemize}

Instead, we categorise the game domain into six regions based on the dynamics for the
purpose of analysis, as shown
in Figure~\ref{fig:gamedomainsLabelled} with labels A, B, C, D, E and F.
Each of these regions is characterised by its boundary conditions. They are:

\begin{itemize}
\item Region A : $T > S+1$
\item Region B : $\tfrac{S+1}{2} < T < S+1$ and $T>2S$
\item Region C : $\tfrac{S+1}{2} < T < 2S$
\item Region D : $2S < T < \tfrac{S+1}{2}$
\item Region E : $S < T < 2S$  and $T < \tfrac{S+1}{2}$
\item Region F : $T < S$
\end{itemize}

Next we prove a lemma that will be used frequently in the proofs.

\begin{lem}
\label{lem:absorptiontime}
Let $\ell_t$ be the length of a d-run at time $t$.
If the length of the run reduces by at least $p$ in expectation for some constant $p>0$, the probability that the run is not deleted in
$\tfrac{\ell_0}{p(1-\varepsilon)}$ generations is at most $e^{-\tfrac{\varepsilon^2\ell_0}{2(1-\varepsilon)}}$ for any $\varepsilon > 0$.
\end{lem}
\begin{proof}
Let $\T$ be the time at which $\ell_t$ becomes $0$ for the first time.
Let $X_t$ be the decrement of the length of the run at time $t$.
Thus
\begin{equation*}
   X_t = \begin{cases}
        0 & \text{if $\ell_t = 0,$}\\
        1 & \text{otherwise.}
        \end{cases}
\end{equation*}

Thus, for $t \le \T$, the decrements are
independent random variables with expectation $\E[X_t] \ge p$ where $X_t \in \{0, 1\}$. Thus
\begin{equation*}
\ell_t = \ell_0 - \sum_{i=1}^t X_i\ .
\end{equation*}

 For any $t \le \T$, using the Chernoff bound, we get
\begin{equation}
\label{e:sumchernoff}
\Pr \Bigg( \sum_{i=1}^t X_i \le pt(1-\varepsilon)\Bigg) \le \exp(-\tfrac{1}{2} \varepsilon^2 pt)  ,
\end{equation}
for any $\varepsilon > 0$. Since $\ell_\T =0$, we have
\begin{equation*}
\sum_{i=1}^\T X_i \ge \ell_0\ .
\end{equation*}

Hence, applying \eqref{e:sumchernoff} for $\T$ steps, we get
\begin{equation*}
\Pr \bigg( \T \ge \dfrac{\ell_0}{p(1-\varepsilon)}\bigg) \le \exp (-\tfrac{1}{2} \varepsilon^2 p\T) \le  \exp\left(-\tfrac{1}{2} \varepsilon^2 p\tfrac{\ell_0}{p(1-\varepsilon)}\right) =   \exp\left(-\tfrac{\varepsilon^2\ell_0}{2(1-\varepsilon)}\right) \ .
\end{equation*}

\end{proof}

Next we explore some properties of the initial random configuration.

\subsubsection{Properties of the initial configuration}

In this study, the initial configuration is assumed to  be generated randomly as follows:
each vertex on the cycle is independently assigned as a cooperator with constant probability
$p_\textrm{c}$ and as a defector with probability $p_\textrm{d} = 1-p_\textrm{c}$ at the
beginning of the game. Hence, at the beginning of the game, the expected
number of cooperators present on the cycle is $np_\textrm{c}$. So, it is reasonable
to assume $p_\textrm{c} = \Omega(1/n)$, so that there will be some cooperators at the start.

\begin{lem}
\label{lem:barrierpresence}
The probability that there is no $k_c$-run, where $k \ge 1$ is a constant,
on the cycle at the beginning of the game is at most
$e^{-p_\textrm{c}^k \lfloor n/k\rfloor}$.
\end{lem}
\begin{proof}
Let $P_k$ be the probability that $k$-consecutive vertices
are not all cooperators. Hence
\begin{equation*}
P_k = (1-p_\textrm{c}^k)\ .
\end{equation*}

There are $\lfloor n/k\rfloor$
disjoint $k$-segments on the cycle. As each vertex
is assigned the initial state independently, these segments are independent.
Hence, the probability that none of those segments is all-cooperators is
\begin{equation*}
P_k^{\lfloor n/k \rfloor} = (1-p_\textrm{c}^k)^{\lfloor n/k\rfloor} \le  e^{-p_\textrm{c}^k \lfloor n/k\rfloor} ,
\end{equation*}
and the lemma is proved.
\end{proof}

\begin{lem}
\label{lem:maxcrun}
The following statement holds with high probability. The longest
c-run generated in the initial configuration is of length $\lambda \log n$ for any $\lambda > 0$ such that $\lambda \log (1/p_\textrm{c}) > 1$.
\end{lem}
\begin{proof}

Let $r_\ell$ be the number of c-runs of length $\ell$. So,
\begin{equation*}
 \E[r_\ell] = n {p_\textrm{c}}^{\ell}\ .
\end{equation*}

If $\ell = \lambda\log n$, then
\begin{equation*}
 \E[r_\ell]\ =\ n  {p_\textrm{c}}^{\ell}\ =\ n  {p_\textrm{c}}^{\lambda  \log n}\ =\ n^{1-\lambda  \log (1/p_\textrm{c})} \ .
\end{equation*}

So,
\begin{equation*}
\Pr ( r_\ell \neq 0) \le n^{1-\lambda  \log (1/p_\textrm{c})} \ .
\end{equation*}

Since $n^{1-\lambda  \log (1/p_\textrm{c})}  \to 0$ as  $n \to \infty \,$ for
 $\lambda \log (1/p_\textrm{c}) > 1$,
we do not expect to see any runs of length  $\lambda\log n$ or longer and the lemma is proved.
\end{proof}

\begin{lem}
\label{lem:maxdrun}
The following statement holds with high probability. The longest
d-run generated in the initial configuration is of length $\lambda \log n$ for any $\lambda > 0$ such that $\lambda \log (1/p_\textrm{d}) > 1$.
\end{lem}

\begin{proof}
The proof is entirely similar to that of Lemma~\ref{lem:maxcrun}, so will be omitted.
\end{proof}

\begin{lem}
\label{lem:maxLengthDE}
Consider a segment on the cycle having d-runs of any length separated by $1_c$-runs.
Such a segment must have a c-run of length greater than 1 at each end. Now, the
following statement about this segment holds with high
probability. The length of the longest such segment in the initial configuration  is $O(\log n)$.
\end{lem}
\begin{proof}
\newcommand{\M}{$\mathcal{M}$}
The length of the segment can  be calculated using a simple random walk. Suppose the vertices
are assigned as cooperators (or ones) or defectors (or zeroes) from $v_0$ to $v_{n-1}$.
At any time $t$, there can be four combinations of present and future
assignments. They are: 00, 01, 10, and 11. Consider these as four states of
a Markov chain \M, and denote these states by $s_0, s_1, s_2,$ and $s_3$ respectively.
The transition probabilities for these states are shown in
Figure~\ref{fig:initialConfig}.

\begin{figure}[ht]
\centering
\subfigure{

\xymatrix{
&*+[o][F] {s_3}   \ar[dl]_{1-p_\textrm{c}} \ar@(ul,ur)[]^{p_\textrm{c}}  & \\
*+[o][F-]{s_2}\ar@/^/[rr]^{p_\textrm{c}}\ar[dr]_{1-p_\textrm{c}} & & *+[o][F-]{s_1} \ar@/^/[ll]^{1-p_\textrm{c}}  \ar[ul]_{p_\textrm{c}} \\
&  *+[o][F-]{s_0}  \ar[ur]_{p_\textrm{c}} \ar@(dl,dr)[]_{1-p_\textrm{c}} &  }
}
\caption{ A random walk representing the generation of initial configuration,
modelled as a four-state Markov chain. The edges of the state diagram are
labelled with transition probabilities.}
\label{fig:initialConfig}
\end{figure}
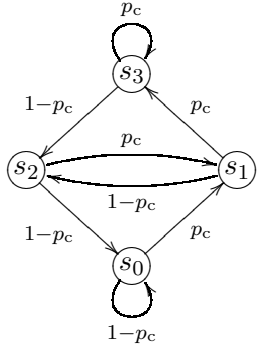

We are interested to find the total length of the d-runs connected by
$1_\textrm{c}$-runs. Thus, we need to find the time to return to $s_3$, or equivalently, the time to go from $s_2$ to $s_3$. This time is exactly
the length of the d-runs separated by singleton cooperators.

Clearly, from any other states of \M, there is a probability at least $p_c^2$ of reaching $s_3$ in two steps.
Hence the probability that $s_3$ will not be reached in $2t$ steps is at most $p_c^{2t}$.
Thus the probability that there exists any such run of length $\lambda \log n$, for any $\lambda > 0$, is at most
\begin{equation*}
 np_\textrm{c}^{\lambda \log n}  < \frac{1}{n}, \mbox{ if } \ \   \lambda > \dfrac{2}{\log(1/p_\textrm{c})}\ .
\end{equation*}

Hence, if $\lambda > \tfrac{2}{\log(1/p_\textrm{c})}$, the probability of finding this special configuration
 of length $\lambda \log n$ tends to 0 as $n \to \infty.$ Thus, with high probability, the maximum length of the segment is $O(\log n)$.
\end{proof}

\begin{lem}
\label{lem:maxLengthAlternating}
The following statement holds with high probability. The longest sequence of alternating
1's and 0's in the initial configuration is at most $2\lambda \log n$, for any $\lambda > 0$ such that
$\lambda \big( \log (1/p_\textrm{c}) + \log (1/p_\textrm{d})\big) > 1$.
\end{lem}
\begin{proof}

Let $r_\ell$ be the number of occurrences of a chain of length $\ell$ having alternating
1's and 0's. Suppose $\ell$ is even,
without loss of generality. Then we have
\begin{equation*}
\E[r_\ell] = n{p_\textrm{c}}^{\tfrac{\ell}{2}}{p_\textrm{d}}^{\tfrac{\ell}{2}}\ .
\end{equation*}
If $\ell =2\lambda \log n$, then
\begin{equation*}
\E[r_\ell]\ =\ n  {p_\textrm{c}}^{\lambda \log n} {p_\textrm{d}}^{\lambda \log n} \ = \ n^{1-\lambda  \log (1/p_\textrm{c}) - \lambda  \log (1/p_\textrm{d})}\ .
\end{equation*}
So, if $\lambda \big( \log (1/p_\textrm{c}) + \log (1/p_\textrm{d}) \big) > 1$,  we have
\begin{equation*}
\Pr ( r_\ell \neq 0) \le \ n^{1-\lambda  \log (1/p_\textrm{c}) -\lambda  \log (1/p_\textrm{d}) } \to 0 \mbox{ as } n \to \infty.\qedhere
\end{equation*}
\end{proof}

\subsubsection{Emergence of defection}
\label{sec:A}

In this section we prove that the defection emerges fast for the games in the region labelled A in Figure~\ref{fig:gamedomainsLabelled}.
This region, having $S+1 < T$, covers the whole PD domain
and half the domain of SH and SG. The table below shows which of the
switching probabilities are zero for the vertex $v_i$.

\smallskip
\begin{center}
\begin{tabular}{|c|c|l|l|l|l|}
  \hline
 \multirow{2}{*}{$\CS_{i}$}  \Tgap & \multirow{2}{*}{$\CS_{i+1}$}& \multicolumn{4}{|c|}{$\CS_{i-1}\CS_{i+2}$} \\
\cline{3-6}
  &  & 00 & 01 & 10 & 11\\
\hline
    $0$ & $1$ & $0$  & $0$  & $0$ & $0$  \\
   \hline
    $1$ & $0$ & $+$ & $+$ & $+$ & $+$  \\
  \hline
\end{tabular}
\end{center}
\smallskip

Here, no defector will ever become a cooperator, but cooperators
can  become defectors. So, the game converges to the all-defect state fast. The following lemmas prove this.

\begin{lem}
\label{lem:fastconvA}
Suppose $\R_\textrm{c}$ is the longest c-run in the cycle when the game is started and
let $\ell(\R_\textrm{c}) =\gamma$.
Provided $S+1 < T$, probability that all-defect state is not reached in
$\frac{\gamma}{\bar{b}(1-\varepsilon)}$
generations is at most $e^{-\tfrac{\varepsilon^2\gamma}{2(1-\varepsilon)}}$,  for any $\varepsilon >0$.
\end{lem}
\begin{proof}
In synchronous updating, each vertex updates its strategy at the end of every generation. When $S+1 < T$,
the only vertices that imitate their neighbours are the cooperators at either end of a c-run.
(Note that a $1_c$-run has only one such cooperator.) A cooperator at this position chooses a defector for imitation
with probability $\tfrac{1}{2}$. Let $p$ be defined to be the minimum of
the four possible switching probabilities (see Figure~\ref{fig:switchingprob2}). It can be easily verified that we have
\begin{equation*}
p = \min\{ \bar{a}, \bar{b}, \bar{c}, \bar{d} \} = \bar{b}\ .
\end{equation*}

Then, an $\ell_\textrm{c}$-run $(2 \le \ell < n)$ reduces in length at either end with probability
at least $\tfrac{1}{2}p$ and a $1_c$-run is deleted with probability at least $p$, in every
generation. Thus, the expected decrease
in the length of any c-run is at least $\bar{b}$.
The time it takes for the longest c-run to be deleted completely is precisely the absorption
time, since shorter runs are deleted faster.  The result then follows from
Lemma~\ref{lem:absorptiontime}.
\end{proof}

%

In the worst case, the length of the longest c-run ($\gamma$) can be $n$. This means
all players cooperate at the beginning,  which is an absorbing state. Hence,
the all-defect state is never reached.
But, if $\gamma = n-1$, in view of Lemma~\ref{lem:fastconvA}, it takes
$\tfrac{(n-1)}{\bar{b}(1-\varepsilon)} \approx \tfrac{n}{\bar{b}}$ generations for the convergence to defection.
However, it was shown in Lemma~\ref{lem:maxdrun} that, when $p_\textrm{c} < 1$, the length of the longest c-run is $O(\log n)$.
Theorem~\ref{thm:fastconvA} combines all these results.

\bigskip
\noindent \textbf{
Proof of Theorem ~\ref{thm:fastconvA}: }
By Lemma~\ref{lem:maxdrun}, the longest c-run present on the cycle
at the beginning of the game is of length $\lambda \log n$ w.h.p for any
$\lambda > 0$ such that $\lambda \log (1/p_\textrm{c}) > 1$.
Then, by Lemma~\ref{lem:fastconvA}, the probability that the steady state
has not been reached in
\begin{equation*}
\dfrac{\lambda \log n }{\bar{b}(1- \varepsilon)}  = O(\log n)
\end{equation*}
generations is at most $n^{-\tfrac{\varepsilon^2\lambda}{2(1-\varepsilon)}}$ for any
$\varepsilon > 0$, where $\bar{b}=\tfrac{T-S-1}{2\alpha}\ $ and $\alpha =  \max\{T, 1\}- \min\{S,0\}$. For a suitable value for $\lambda$, this probability tends to 0 as $n \to \infty$ which completes the proof.
\hfill $\Box$

\subsubsection{Emergence of cooperation}
\label{sec:emergenceofcoop}

In this section, we prove that cooperation emerges fast
in regions B, C, D, E, and F. Before analysing these,
let us first look at
some common characteristics shown by the regions B, C, D and E.
Figure~\ref{fig:dynamicsBCDE} shows which of the switching
probabilities are zero for each of these regions. Note, from Figure~\ref{fig:gamedomainsLabelled}, that the regions B, C, D and E satisfy $T < S+1$. Hence, Lemma~\ref{lem:barrier} holds for these regions.

\begin{figure}
\centering
\subfigure[Region B]{
\label{fig:dynamicsB}
\begin{tabular}{|c|c|l|l|l|l|}
  \hline
 \multirow{2}{*}{$\CS_{i}$}  \Tgap & \multirow{2}{*}{$\CS_{i+1}$}& \multicolumn{4}{|c|}{$\CS_{i-1}\CS_{i+2}$} \\
\cline{3-6}
  &  & 00 & 01 & 10 & 11\\
\hline
    $0$ & $1$ & $0$  & $+$  & $0$ & $0$  \\
   \hline
    $1$ & $0$ & $+$ & $+$ & $0$ & $+$  \\
  \hline
\end{tabular}
}
\hspace{1cm}
\subfigure[Region C]{
\label{fig:dynamicsD}
\begin{tabular}{|c|c|l|l|l|l|}
  \hline
 \multirow{2}{*}{$\CS_{i}$}  \Tgap & \multirow{2}{*}{$\CS_{i+1}$}& \multicolumn{4}{|c|}{$\CS_{i-1}\CS_{i+2}$} \\
\cline{3-6}
  &  & 00 & 01 & 10 & 11\\
\hline
    $0$ & $1$ & $+$  & $+$  & $0$ & $0$  \\
   \hline
    $1$ & $0$ & $0$ & $+$ & $0$ & $+$  \\
  \hline
\end{tabular}
}\\
\subfigure[Region D]{
\label{fig:dynamicsC}
\begin{tabular}{|c|c|l|l|l|l|}
  \hline
 \multirow{2}{*}{$\CS_{i}$}  \Tgap & \multirow{2}{*}{$\CS_{i+1}$}& \multicolumn{4}{|c|}{$\CS_{i-1}\CS_{i+2}$} \\
\cline{3-6}
  &  & 00 & 01 & 10 & 11\\
\hline
    $0$ & $1$ & $0$  & $+$  & $0$ & $+$  \\
   \hline
    $1$ & $0$ & $+$ & $+$ & $0$ & $0$  \\
  \hline
\end{tabular}
}
\hspace{1cm}
\subfigure[Region E]{
\label{fig:dynamicsE}
\begin{tabular}{|c|c|l|l|l|l|}
  \hline
 \multirow{2}{*}{$\CS_{i}$}  \Tgap & \multirow{2}{*}{$\CS_{i+1}$}& \multicolumn{4}{|c|}{$\CS_{i-1}\CS_{i+2}$} \\
\cline{3-6}
  &  & 00 & 01 & 10 & 11\\
\hline
    $0$ & $1$ & $+$  & $+$  & $0$ & $+$  \\
   \hline
    $1$ & $0$ & $0$ & $+$ & $0$ & $0$  \\
  \hline
\end{tabular}
}
\caption[Game dynamics  for B, C, D and E]{Switching probabilities for B, C, D and E.}
\label{fig:dynamicsBCDE}
\end{figure}

\begin{lem}
\label{lem:barrier}
In  B, C, D and E, if a c-run $\R_\textrm{c}$ of length at least $2$
is adjacent to d-runs of length at least $2$ at each end, the c-run grows in
length or remains as it is. The expected growth of the run in one generation is $b$.
\end{lem}
\begin{proof}

The proof is based on the observation that  the switching probability
is positive for $0[0]11$ and zero for $1[1]00$  in the regions
in question.

Now, suppose $\R_\textrm{c} = [i,j]$ such that  $\ell(\R_\textrm{c}) \ge 2$.
Suppose $\R_\textrm{c}$ is bordered by at least two
defectors on both sides, so we have $\star 001100 \star$. The
 vertices of the outer-rim  edges of $R_\textrm{c}$, namely $v_{i-1}, v_i, v_{j}$ and $v_{j+1}$,
are in states  $0, 1, 1$ and $0$ respectively.
Possible changes to these vertices after a generation are:

\begin{itemize}
  \item $v_{i-1}$, which is a defector itself, has another defector on its left and a cooperator on its right. Hence, switching can  happen only if  $v_{i-1}$ tries to imitate from its right.  As Figure~\ref{fig:switchingprob2}
  indicates, this happens with probability $\tfrac{1}{2}b$ (as the right neighbour is chosen with probability $\tfrac{1}{2}$
  and the actual switching happens independently with probability $b$ which is non-zero according to Figure~\ref{fig:dynamicsBCDE}).
  This makes $\R_\textrm{c}$ longer in length by $1$.
  \item $v_i$, which is a cooperator itself, has another cooperator on its right and  a defector on its left. Hence an effective
  imitation can happen only when it copies from its left neighbour. But, as shown in  Figure~\ref{fig:dynamicsBCDE}, the
  probability of switching to defection in this scenario (i.e. $1[1]00$) is zero. Hence, this vertex will not change its strategy.
  \item By symmetry, $v_j$ does the same as $v_i$.
  \item By symmetry, $v_{j+1}$ does the same as $v_{i-1}$.
\end{itemize}

Hence, $\R_\textrm{c}$ grows in length  with some non-zero probability.
The expected growth of $\R_\textrm{c}$ at either end is $\tfrac{1}{2}b$ in one generation.
Hence the total expected growth of the run equals $b$.
\end{proof}

\begin{rem}
 Lemma~\ref{lem:barrier} implies that an $\ell_\textrm{c}$-run ($\ell \ge 2$) bordered by at least
2 defectors is hard to eliminate in B, C, D and E. We say ``hard" because, as we will see later,
 one such configuration can be eliminated by another in some regions. By combining
 this observation with the dynamics of each region separately, we later establish the conditions determining a \emph{barrier} for each region.
\end{rem}

Now we investigate the dynamics of a singleton cooperator having at least two
defectors at its either end.

\begin{lem}
\label{lem:growingSingletonPlus}
A singleton cooperator ($1_\textrm{c}$-run) bordered on both sides by at least two defectors (i.e.\ $\star00100\star$)
can grow up to length 3 in C and E, whereas it is deleted in B and D.
\end{lem}
\begin{proof}
Let $R_\textrm{c} =[i,i]$ be a $1_\textrm{c}$-run and let
$\CS_{i-2}= \CS_{i-1} = \CS_{i+1} = \CS_{i+2} = 0$. Here, $v_i$ might imitate both its neighbours
and has $0[1]00$ in both cases. As the comparison tables
in Figure~\ref{fig:growingSingeltonPlus} show, the switching probability for
 $0[1]00$ is positive in B and D, and zero in E and C. Hence, $v_i$ might become a
 defector in B and D, but will remain as a cooperator in E and C.

\begin{figure}[ht!]
\centering
\subfigure[B and D]{
\begin{tabular}{|c|c|c|}
  \hline
 \multirow{2}{*}{$\CS_{i}$}  \Tgap & \multirow{2}{*}{$\CS_{i+1}$}&  $\CS_{i-1}\CS_{i+2}$ \\
\cline{3-3}
  &  & 00 \\
\hline
    $0$ & $1$ & $0$   \\
  \hline
    $1$ & $0$ & $+$    \\
  \hline
\end{tabular}
}
\hspace{1cm}
\subfigure[C and E]{
\begin{tabular}{|c|c|c|}
  \hline
 \multirow{2}{*}{$\CS_{i}$}  \Tgap & \multirow{2}{*}{$\CS_{i+1}$}&  $\CS_{i-1}\CS_{i+2}$ \\
\cline{3-3}
  &  & 00 \\
\hline
    $0$ & $1$ & $+$   \\
   \hline
    $1$ & $0$ & $0$    \\
  \hline
\end{tabular}
}
\caption{The switching probabilities related to singleton cooperators}
\label{fig:growingSingeltonPlus}
\end{figure}

Next, let us see what happens to the neighbours of $R_\textrm{c}$, namely $v_{i-1}$ and $v_{i+1}$.
It can be easily verified that both  $v_{i-1}$ and $v_{i+1}$ have a defector at one side and the cooperator $v_i$ at the other side. So, the
switching can happen only when these vertices copy from $v_i$. In that case, the defectors appear as
$0[0]10$ which has a zero switching probability in B and D and a non-zero switching probability in C and E.

To sum up, in B and D, while $v_i$ switches to defection, its defector neighbours
remain unchanged, hence $R_\textrm{c}$ is deleted. But, in C and E, while $v_i$ remains as a cooperator,
its neighbours can become cooperators too, potentially increasing the length of the $1_c$-run up to 3.
\end{proof}

A singleton defector having longer c-run neighbours has the potential to grow in regions B and C, as
the following lemma shows.

\begin{lem}
\label{lem:growing1drun}
A singleton defector ($1_\textrm{d}$-run) bordered on both sides by at least two cooperators (i.e.\ $\star 11011 \star$)
can grow up to length 3 in B and C, whereas it is deleted in D and E.
\end{lem}
\begin{proof}
In this case, the defector in the middle could imitate from both its neighbours and
has the neighbourhood of $1[0]11$ on either side, and the adjacent cooperators can
copy only from the middle defector and have $1[1]01$. The dynamics
for these two cases are compared in Figure~\ref{fig:growingSingeltonMinus}.

\begin{figure}[ht!]
\centering
\subfigure[B and C]{
\begin{tabular}{|c|c|c|}
  \hline
 \multirow{2}{*}{$\CS_{i}$}  \Tgap & \multirow{2}{*}{$\CS_{i+1}$}&  $\CS_{i-1}\CS_{i+2}$ \\
\cline{3-3}
  &  & 11 \\
\hline
    $0$ & $1$ & $0$   \\
   \hline
    $1$ & $0$ & $+$    \\
  \hline
\end{tabular}
}
\hspace{1cm}
\subfigure[E and D]{
\begin{tabular}{|c|c|c|}
  \hline
 \multirow{2}{*}{$\CS_{i}$}  \Tgap & \multirow{2}{*}{$\CS_{i+1}$}&  $\CS_{i-1}\CS_{i+2}$ \\
\cline{3-3}
  &  & 11 \\
\hline
    $0$ & $1$ & $+$   \\
   \hline
    $1$ & $0$ & $0$    \\
  \hline
\end{tabular}
}
\caption{The switching probabilities related to singleton defectors}
\label{fig:growingSingeltonMinus}
\end{figure}

As Figure~\ref{fig:growingSingeltonMinus} shows, in E and D, the switching probability is positive
for $1[0]11$ and zero for $1[1]01$. Consequently, in $\star 11011 \star$,
the defector might become a cooperator while its neigbouring cooperators remain  unchanged. Hence,
the $1_d$-run is deleted in E and D. But, in the case of B and C, the opposite is true: the
middle defector will remain as it is while its cooperator neighbours switch to defection. Thus,
the singleton defector might become a $2_\textrm{d}$-run or a $3_\textrm{d}$-run as claimed.
\end{proof}

Let us now analyse each region separately.

\bigskip
\noindent\textbf{Region B ($\tfrac{S+1}{2} < T < S+1$ and $T>2S$)}
\label{sec:B}

Figure~\ref{fig:gamedomainsLabelled} shows this region with label
B. Figure~\ref{fig:dynamicsB} shows which cases have zero and
non-zero switching probabilities.
In this section we prove that cooperation
evolves in linear time in this region. Analysing the actual imitation process to prove this is complicated.
Fortunately, we can use a simplified
model  for this purpose. The following lemma forms the basis for the simplification of the process.

\begin{lem}
\label{lem:barrierB}
An $\ell_\textrm{c}$-run ($\ell \ge 4$) is a barrier in region B, whereas a c-run shorter
than 4 might be deleted through a sequence of updates.
\end{lem}
\begin{proof}

In region B,
\begin{description}
  \item [a $1_\textrm{c}$-run is always deleted:~] Lemma~\ref{lem:growingSingletonPlus} shows
  that a $1_\textrm{c}$-run is deleted if it has two defectors on both sides.
  Now, when  a $1_\textrm{c}$-run  is adjacent to a $1_\textrm{d}$-run on either side, as in $\star 10[1]01 \star$,
  it is readily verified that the  $1_\textrm{c}$-run will be turned into a defector whereas the adjacent
  defectors remain unchanged. Thus,  a $1_\textrm{c}$-run cannot grow if
it is adjacent to singleton defectors at both ends. Obviously this would be the case when  a $1_\textrm{c}$-run has
  a singleton defector on one side and at least two defectors on the other side.

It is noteworthy that if all the cooperators on the cycle exist
as $1_\textrm{c}$-runs, the game converges to all-defect, since
singleton cooperators can never survive.

  \item [a $2_\textrm{c}$-run can grow or be deleted:~] A $2_\textrm{c}$-run grows if it is
  bordered by at least two defectors as shown in Lemma~\ref{lem:barrier}. Now, when a $2_\textrm{c}$-run
  is adjacent to $1_\textrm{d}$-runs on both sides, either of the $1_\textrm{d}$-runs can grow,
  deleting a cooperator in the $2_\textrm{c}$-run, as illustrated in Lemma~\ref{lem:growing1drun}, and
  removing the $2_\textrm{c}$-run completely. Thus, when a $2_\textrm{c}$-run is adjacent to a $1_\textrm{d}$-run on one side and at least two defectors on the other, it has the possibility of growing or
reducing to a $1_\textrm{c}$-run which is subsequently deleted.

An interesting case shows that even a $2_\textrm{c}$ run bordered  by at least two defectors can be
removed if it is adjacent to another $2_\textrm{c}$-run. Consider
the configuration $0011001100$. This might first become $0011101100$, then become $001100010$,
and finally become $001100000$, deleting one of the $2_\textrm{c}$-runs completely.

  \item [a $3_\textrm{c}$-run can grow or be deleted:~] Like the pair of cooperators, although a $3_\textrm{c}$-run
  can grow when it is bordered by two defectors at either end, there is a possibility of it being deleted if it
has singleton defectors at  both  ends.

\item [an $\ell_\textrm{c}$-run ($\ell \ge 4$) can never be deleted:~] Even
if there are singleton defectors at either end of a $4_\textrm{c}$-run , the run's length will be reduced to 2
in the worst case. The resulting $2_\textrm{c}$-run will then be  bordered by two defectors, hence will start growing
again, as shown in Lemma~\ref{lem:barrier}. Obviously, longer runs are more stable and cannot
be deleted.

\end{description}

In short, the key observations are: $1_\textrm{c}$-runs are always deleted;
$2_\textrm{c}$-runs and $3_\textrm{c}$-runs might grow or be deleted; and,
runs of length 4 or more can never be deleted. Hence a run of length 4 or more is a \emph{barrier}.
\end{proof}

Clearly, the worst case for the evolution of cooperation is when there is only one
minimal barrier (a  $4_\textrm{c}$-run) at the
start of the game. We use this worst case scenario to determine the absorption time. But, we need to address an issue before doing that.
Although  the time calculated in this way gives the worst case for a c-run to grow until the
all-cooperate state is reached, it might not include the overhead time required if there are many runs.
There are two types of such overheads: the time spent on handling short
runs that can become a \emph{barrier} or be deleted, and the time spent on merging two c-runs. We first calculate the worst case time for these
overheads in Lemma~\ref{lem:shortRunsB} and \ref{lem:mergingTimeB} respectively.

\begin{lem}
\label{lem:shortRunsB}
Let $\phi_s$ denote the expected time it takes for a $1_\textrm{c},2_\textrm{c},3_\textrm{c}$-run to be
deleted or become a \emph{barrier} (a run of length 4). Then $\phi_s = O(1)$.
\end{lem}
\begin{proof}
Recall that a singleton cooperator can never grow. According to Figure~\ref{fig:switchingprob2},
it is deleted with probability
\begin{itemize}
    \item $\bar{a}$ if it is in the form $00[1]00$,
    \item $\bar{c}$ if it is in the form $10[1]01$, and
    \item $\tfrac{1}{2}(\bar{a} + \bar{c})$ if it is in the form $10[1]00$.
\end{itemize}

We also have $\bar{a} < \bar{c}$. Hence, the worst expected time it takes for a $1_\textrm{c}$
to be removed is $1/\bar{a}$,  since it has the geometric distribution with probability
of success of $\bar{a}$. Now it is enough to determine the time it takes for a $2_\textrm{c}$-run
or a $3_\textrm{c}$-run to become a $1_\textrm{c}$-run or a $4_\textrm{c}$-run, as we know that
$1_\textrm{c}$'s are removed in constant time. For calculating this,
consider a random walk on the number of cooperators, i.e.\ \{1,2,3,4\}. Then we need to show that
the walk reaches 1 or 4 in constant time irrespective of where it starts. Next we determine the
corresponding probabilities of this process.

Suppose the current position of the random walk is 2 or 3, i.e.\ there
is a $2_\textrm{c}$-run or a $3_\textrm{c}$-run respectively. A $2_\textrm{c}$-run
or a $3_\textrm{c}$-run can exist in three forms:

\begin{itemize}
    \item $001100$ or $0011100$. In this case, the probability of going to the right is
$b - \tfrac{1}{4}b^2$ and the probability of going to the left is 0.
    \item $001101$ or $0011101$. In this case, the probability of going to the right is
$\tfrac{1}{2}b\bigl(1-\tfrac{1}{2}\bar{d}\bigr)$ and the probability of going to the left
is $\tfrac{1}{2}\bar{d}\bigl(1-\tfrac{1}{2}b\bigr)$.
    \item $101101$ or  $1011101$. In this case, the probability of going to the right is
$0$ and the probability of going to the left is $ \bar{d}- \tfrac{1}{4}\bar{d}^2$.
\end{itemize}

We want an upper bound on the time it takes to reach state $1$ or $4$.
It takes longest to reach $4$ if the probability of moving to the right
is the minimum non-null probability, $r$ say. If the probability of moving to
right is zero, the walk never reaches $4$ and is absorbed at $0$. We have
\begin{equation*}
r = \min \bigl\{b- \tfrac{1}{4}b^2, \tfrac{1}{2}b\bigr(1-\tfrac{1}{2}\bar{d}\bigl)\bigr\}\ .
\end{equation*}

Similarly, it takes longest to reach $1$ when the probability $q$ of moving left from $2$ or $3$ is
\begin{equation*}
q = \min \bigl\{\bar{d}- \tfrac{1}{4}\bar{d}^2, \tfrac{1}{2}\bar{d}\bigr(1-\tfrac{1}{2}b\bigl)\bigr\}\ .
\end{equation*}

So we have a Gambler's Ruin problem with absorbing barriers at $1$ and $4$.
Using the standard results when $r+q=1$ (see, for example, Feller\cite[p.~345]{FellerVolI}),
if the game is started from state $\ell \in\{2,3\}$, the expected duration of the process is
\begin{equation*}
   D_\ell = \begin{cases}
        \dfrac{\ell}{q-r} - \dfrac{4}{q-r}\dfrac{1-(q/r)^\ell} {1-(q/r)^4} & \text{if $r \neq q$,}\\
       \ell(4-\ell) & \text{if $r = q .$}
        \end{cases}
\end{equation*}

Let $D_{\max} =\max \{D_2,D_3\}$. Clearly $D_{max}$ is constant as it only involves
the constants $q$ and $r$ which, in turn, can be expressed in terms of the constants $S$ and $T$.
There is a self-loop at 2 and 3 with constant probability, at most $1-(r+q)$, but this slows
down the random walk only by a constant factor. Thus it takes only constant time to reach $1$ or $4$.
We showed earlier that it takes constant time to go from $1$ to $0$.
Hence the total expected time is constant.
\end{proof}

The other overhead of having more than one run is the time required to merge them. Two barriers merge together when
they are separated by two defectors and both defectors switch to cooperation simultaneously. As
the switching of both defectors happens independently with some probability, there is a
possibility that only one of the two defectors switches to cooperation whereby a singleton defector
is created between the two c-runs. Then, as shown in Lemma~\ref{lem:growing1drun},
the singleton defector can grow up to length $3$. And then the barriers start growing again. This is repeated until
the runs are merged. The following lemma proves that the worst case merging time for two
c-runs of length at least 4 is a constant.

\begin{lem}
\label{lem:mergingTimeB}
Let $\phi_m$ denote the expected time it takes for two barriers (i.e.\ two c-runs of length at least 4)
separated by 1 or 2 defectors to merge together. Then $\phi_m = O(1)$.
\end{lem}
\begin{proof}
The merging process can be modelled as a simple absorbing Markov chain with states
$M_i$ ($ 0 \le i \le 3$) where $i$ denote the number of defectors in between the
barriers. $M_0$
is the absorbing state and all other states are transient. Using Figure~\ref{fig:switchingprob2},
we can determine the transition probabilities for this Markov chain. The corresponding transition matrix in canonical form is

\smallskip
\begin{displaymath}
\kbordermatrix{
   & M_1 & M_2 & M_3 & M_0 \\
M_1 & (1-\bar{d}/2)^2 & \bar{d}(1-\bar{d}/2) & \bar{d}^2/4 & 0 \\
M_2 & b(1-b/2) & (1-b/2)^2 & 0 & b^2/4 \\
M_3 & b^2/4 & b(1-b/2) & (1-b/2)^2 & 0\\
M_0 & 0 & 0  & 0 & 1}\ .
\end{displaymath}
\smallskip

Note that all transition probabilities are constants. Hence, using standard methods,
absorption time can be  calculated. Let $\tau_i$ be the absorption time when the chain
starts at $M_i$. Then we get

\smallskip
\begin{equation*}
\begin{bmatrix}
\tau_1\\
\tau_2\\
\tau_3
\end{bmatrix}
=
\begin{bmatrix}
\dfrac {16 b-8{b}^{2}+{b}^{3}+16 \bar{d}-4 b\bar{d}-b{\bar{d}}^{2}}{\bar{d}{b}^{2}(4-b-\bar{d}) } \ \\[0.5ex]
\dfrac {2(8b-6 b^2+b^3+8 \bar{d}-2 b \bar{d}-b d^2)}{\bar{d} b^2 (4-b-\bar{d}) }\ \\[0.5ex]
\dfrac {16 b-12 b^2+3  b^3+16 \bar{d}-3 b \bar{d}^2}{\bar{d} b^2 ( 4-b-\bar{d}) }
\end{bmatrix}\ .
\end{equation*}
\smallskip

The worst case merging time $\phi_m = \max \{\tau_1, \tau_2, \tau_3\}$.
Since the time calculation involves only constants, $\phi_m$ is constant, and the lemma is proved.
\end{proof}

Now we will assume that when the game is started there is a $4_\textrm{c}$-run and
an $(n-4)_\textrm{d}$-run on the cycle. The following lemma proves that it takes linear
time for the $4_\textrm{c}$-run to grow up to length $n-2$.

\begin{lem}
\label{lem:growingTimeB}
If the game is started with a $4_\textrm{c}$-run and
an $(n-4)_\textrm{d}$-run on the cycle, the expected
time it takes for cooperation to spread to $n-2$ positions, denoted by $T_g$, is $\Theta(n)$ w.h.p.
\end{lem}
\begin{proof}
As Lemma~\ref{lem:barrier} shows, an $\ell_\textrm{c}$-run ($\ell \ge 2$) continues to grow as
long as it is bordered by at least two defectors. As we assume that the game is started with only
a $4_\textrm{c}$-run on the cycle,  this run can grow up to length $n-2$.

Now, let $T_i$ be an upper bound on the number of steps taken to go from $i$ cooperators to $(i+1)$ cooperators
on the cycle. Then, from
Figure~\ref{fig:switchingprob2}, the adjacent defectors at both
ends of the c-run switch to cooperation with probability $\tfrac{1}{2}b$. Hence, the probability of
increasing the length by at least 1 is $1-\bigl(1-\tfrac{1}{2}b\bigr)^2 $. Let us denote this probability by $p_1$.  Thus   we have
\begin{equation*}
 \Pr (T_i = t) = (1-p_1)^{t-1}p_1\  .
\end{equation*}

$T_i$ has a geometric distribution with parameter $p_1$. Hence, the total expected time $T_g$ is
\begin{equation*}
 \E[T_g] = \sum_{i=4}^{n-3} \E[T_i] = \dfrac{(n-6)}{p_1} = O(n)\  .
\end{equation*}

Let us now bound the probability of getting large deviations
from the mean $\E[T_g]$. From the definition of $T_i$, we obtain
\begin{equation*}
\Pr (T_i > t)  \le (1-p_1)^{t} \le e^{-p_1t} \ .
\end{equation*}

Thus, for $t = \tfrac{\beta  \log n}{p_1}$, we get
\begin{equation*}
\Pr \big(T_i >  \tfrac{\beta  \log n}{p_1}\big)  \le \dfrac{1}{n^\beta}\ .
\end{equation*}

Thus, deviations of size $\tfrac{\beta n \log n}{p_1}$ are unlikely. In other words, $T_i$ lies within the range
$\left[0, \frac{\beta \log n}{p_1}   \right]$ with high probability. Now, define a set of random variables $Y_i$
such that $Y_i = \frac{p_1 T_i}{\beta \log n}$, and let $Y=\sum_{i=4}^{n-3}Y_i$. Then, $Y_i \in [0, 1]$ with high probability, and we have
\begin{equation*}
\E[Y] = \frac{p_1 \E[T_g]}{ \beta \log n} = \dfrac{n-6}{\beta \log n}\ .
\end{equation*}

Since $Y_4, Y_5, \ldots , Y_{n-2}$ are independent random variables taking values in [0,1], we may apply Chernoff-Hoeffding to get
\begin{equation*}
\Pr\bigl(Y \notin (1 \pm \varepsilon) \E[Y] \bigr) \le 2  e^{- \frac{1}{3} \varepsilon^2 \frac{n-6}{\beta \log n}}\ .
\end{equation*}

If $\varepsilon = \frac{3 \beta \log n}{\sqrt{ n-6}}$, the following holds for sufficiently large $n$.
\begin{equation*}
\Pr\left(Y \notin (1 \pm \varepsilon)\E[Y] \right) \le 2 e^{- 3 \beta \log n} = \dfrac{2}{n^{ 3\beta}}\ .
\end{equation*}

It follows immediately that $T_g$ lies within the range  $(1 \pm \varepsilon) \E[T_g]$ with high probability.
Thus we can conclude that $T_g \in \tfrac{(n-6)}{p_1} \pm   O(\sqrt{n}\log n) $, so $T_g = \Theta(n)$ with high probability.
\end{proof}

\bigskip
\noindent \textbf{
Proof of Theorem~\ref{thm:fastconvBandCfinal} (Case - I ):}\ \
In the imitation process, runs just
grow or decrease in length and no runs are ever created. Decreasing in length might mean the
removal of runs. When a d-run is removed, two c-runs are merged and vice versa. Thus, the
worst case absorption time includes the following:

\begin{enumerate}
    \item $T_g$ - the worst case time required for c-runs to grow as much as possible,
	i.e.\ the time taken for a single barrier to become an $(n-2)_\textrm{c}$-run which is $O(n)$ by Lemma~\ref{lem:growingTimeB}.
	\item $T_m$ - the worst case time required for merging c-runs. There can be at most $O(n)$ runs on the
cycle. And the worst case time for merging two c-runs is $O(1)$ by Lemma~\ref{lem:mergingTimeB}.
Hence the total time spent on merging c-runs is $O(n)$.
\item $T_s$ - the worst case time required to handle the c-runs that are not barriers.
This time is spent on growing shorter runs to become barriers or removing them. In
Lemma~\ref{lem:shortRunsB}, it was shown that the time taken to handle one short run
is $O(1)$. Thus, the time to handle all small runs is $O(n)$.
\end{enumerate}

Thus, the worst case absorption time $T= T_g +  T_m + T_s = O(n)$, as claimed.
Finally, it is worth emphasizing the fact that, in the actual process, these three
different types of events happen simultaneously. But, we have added the times
in order to get an upper bound.

What remains is to show that there will be a barrier, a c-run
of length greater than 3, at the beginning
of the game. But it follows from Lemma~\ref{lem:barrierpresence} that
not finding a barrier is exponentially unlikely, so the proof is complete.
\hfill $\Box$

\bigskip
\noindent\textbf{Region C ($\tfrac{S+1}{2} < T < 2S $)}
\medskip
\label{sec:C}

It is easily observed on Figure~\ref{fig:dynamicsBCDE} that the only difference between the
dynamics of region B and C is: singleton cooperators adjacent
to at least two defectors can grow in C, but not in B, as proved in
Lemma~\ref{lem:growingSingletonPlus}. Hence, the characteristics of C can
be summarised as follows:

\begin{itemize}
    \item a $1_\textrm{c}$-run can grow or be deleted.
    \item a $2_\textrm{c}$-run can grow or be deleted.
    \item a $3_\textrm{c}$-run can grow or be deleted
    \item an $\ell_\textrm{c}$-run ($\ell \ge 4$) can never be deleted.
\end{itemize}

Thus,  $\ell_\textrm{c}$-run ($\ell \ge 4$) is a \emph{barrier} in C too.
Due to its similarity to B, the analysis of region B in Section~\ref{sec:B}
will be applicable for C, apart from the time
required to deal with short runs which is determined in
the lemma below.

\begin{lem}
\label{lem:shortRunsD}
Suppose $\phi_s$ is the expected time it takes for a $1_\textrm{c}$, $2_\textrm{c}$, $3_\textrm{c}$-run
to be deleted or become a barrier (a run of length 4). Then, $\phi_s = O(1)$.
\end{lem}
\begin{proof}

Technically speaking, $1_\textrm{c}$-runs, $2_\textrm{c}$-runs and $3_\textrm{c}$-runs
perform a random walk before they become $0_\textrm{c}$-runs or $4_\textrm{c}$-runs. Hence, we consider a
random walk on the number of cooperators, i.e.\ \{0,1,2,3,4\}. Now, it suffices to show
that the walk reaches 0 or 4 in constant time irrespective of where the walk starts.

Suppose the current position of the walk is 1, i.e.\ there is a $1_\textrm{c}$-run.
A $1_\textrm{c}$-run can exist in three forms giving rise to three different cases:
\begin{itemize}
    \item $00[1]00$. In this case, the probability of moving right is
	$a - \tfrac{1}{4}a^2$ and the probability of moving left is 0;
    \item $00[1]01$. In this case, the probability of moving right
	is $\tfrac{1}{2}a\bigl(1-\tfrac{1}{2}\bar{c}\bigr)$ and the probability
of moving left is $\tfrac{1}{2}\bar{c}\bigl(1-\tfrac{1}{2}a\bigr)$;
    \item $10[1]01$. In this case, the probability of moving right is 0 and
the probability of moving left is $\bar{c}$.
\end{itemize}

Next, suppose the current position is 2 or 3, i.e.\ there is $2_\textrm{c}$-run or
$3_\textrm{c}$-run respectively. The probabilities of movement from these states are
the same as for region B. But we give them here for easy reference.
A $2_\textrm{c}$-run or $3_\textrm{c}$-run can exist in three forms:

\begin{itemize}
    \item $001100$ or $0011100$. In this case, the probability of moving right is
	$b - \tfrac{1}{4}b^2$ and the probability of moving left is 0.
    \item $001101$ or $0011101$. In this case, the probability of moving right is
$\tfrac{1}{2}b\left(1-\tfrac{1}{2}\bar{d}\right)$ and the probability of moving
left is $\tfrac{1}{2}\bar{d}\left(1-\tfrac{1}{2}b\right)$.
    \item $101101$ or  $1011101$. In this case, the probability of moving right is
$0$ and the probability of moving left is $\bar{d} - \tfrac{1}{4}\bar{d}^2$.
\end{itemize}

What we want is an upper bound on the time required to reach 0 or 4.
It takes longest to reach 4 if the probability of moving to the right
takes the minimum non-null probability $r$. Hence we get
\begin{equation*}
r = \min \left\{a - \tfrac{1}{4}a^2,\,\tfrac{1}{2}a\left(1-\tfrac{1}{2}\bar{c}\right), \,b - \tfrac{1}{4}b^2, \,\tfrac{1}{2}b\left(1-\tfrac{1}{2}\bar{d}\right)\right\}\ .
\end{equation*}

Similarly, it takes longest to reach 0 when the probability $q$ of moving left is
\begin{equation*}
q = \min \left\{\bar{c}, \,\tfrac{1}{2}\,\bar{c}\left(1-\tfrac{1}{2}a\right), \,\bar{d} - \tfrac{1}{4}\bar{d}^2, \,\tfrac{1}{2}\bar{d}\left(1-\tfrac{1}{2}b\right)\right\}\ .
\end{equation*}

Now, this is simply a Gambler's Ruin problem with absorbing barriers at 0 and 4.
The worst case expected duration of the game can therefore be calculated as in Lemma~\ref{lem:shortRunsB} and shown to be $O(1)$.
\end{proof}

\bigskip
\noindent \textbf{
Proof of Theorem ~\ref{thm:fastconvBandCfinal} (Case - II ): } Recall that the
absorption time
\begin{equation*}
T = T_g + O(n) \phi_m + O(n)\phi_s,
\end{equation*}
where, $T_g$  and $\phi_m$ are the same as for B, and $\phi_s$ is still $O(1)$
for C as proved in Lemma~\ref{lem:shortRunsD}. Hence, $T =O(n)$ w.h.p.
\hfill $\Box$
\bigskip

\bigskip
\noindent\textbf{Regions D  ($2S < T < \tfrac{S+1}{2}$) and E ($S < T < 2S$ and $T < \tfrac{S+1}{2}$)}
\medskip
\label{sec:DandE}

In regions D and E, a $2_\textrm{c}$-run is a  barrier. The following lemma establishes this.

\begin{lem}
\label{lem:barrierDE}
An $\ell_\textrm{c}$-run ($\ell \ge 2$) grows or remains unchanged,
hence is a barrier in  D and E.
\end{lem}
\begin{proof}
Lemma~\ref{lem:barrier} proved that when a c-run of length at least
two is adjacent to d-runs of length at least two, the c-run grows
in length. What is remaining to be shown is,  a c-run of length at least
two is not
deleted or reduced in length when it is adjacent to singleton defectors.
This is a direct result of Lemma~\ref{lem:growing1drun} which
proves that when a $1_\textrm{d}$-run is adjacent to at least two cooperators,
the $1_\textrm{d}$-run cannot grow in length in D and E.
\end{proof}

We will call the part of the cycle  not containing any barriers the \emph{non-barrier}.
A \emph{non-barrier-segment} is a set of vertices between two barriers.
In region C, the non-barrier-segments have only defectors and singleton cooperators.
The length of each non-barrier-segment decreases during the evolution since it
has a barrier  at either end which grows or remains unchanged.
As the update is done synchronously, the length of every non-barrier-segment reduces
in expectation. The all-cooperate state is reached when the lengths of all
non-barrier-segments reach $0$. Obviously, the absorption time is dominated
by the longest non-barrier-segment present at the beginning of the game.

\begin{lem}
\label{lem:fastconvDandE}
Suppose $R_\textrm{cd}$ is the longest non-barrier-segment at the beginning
of the game and let $\gamma$
be its initial length.
In regions D and E,  the probability that the all-cooperate state is not
reached in $\frac{\gamma}{d(1- \varepsilon)}$
generations is at most $e^{-\tfrac{\varepsilon^2\gamma}{2(1-\varepsilon)}}$, for any $\varepsilon > 0$.
\end{lem}

\begin{proof}
Let $\chi$ be the longest non-barrier segment, spanning from $v_i$ to $v_j$.
Let $\sigma_t$ be the length of $\chi$ after $t$ generations.
Hence, the time $t$ at which $\sigma_t$ becomes $0$ is the absorption time.

Let us first determine the expected minimum negative growth of the $\chi$ at its left end, i.e.\
along the edge $\{i-1, i\}$. Lemma~\ref{lem:barrierDE} shows that $\ell_\textrm{c}$-run ($\ell \ge 2$)
never reduces in length. This means that $v_{i-1}$ never switches to defection. So we determine the minimum
probability that $v_{i}$ becomes a cooperator. The switching of $v_{i}$ depends on the status of
$v_{i-2},v_{i-1},v_{i+1}$ and $v_{i+2}$.  Let $\CS_i^t$ denote the status of the vertex $v_i$
after $t$ generations. Using Figure~\ref{fig:dynamicsBCDE}, the following table presents the
possible values for the probability that $v_{i}$  becomes  a cooperator in D and E.

\smallskip
\begin{center}
\begin{tabular}{|c|c|c|c|c|c|}
  \hline
  $\CS_{i-2}^t$ & $\CS_{i-1}^t$ & $\CS_{i}^t$ & $\CS_{i+1}^t$ & $\CS_{i+2}^t$ & $\Pr\bigl[ \CS_{i}^{t+1} = 1\bigl]$ \\
 \hline
  1 & 1 & 0 & 0 & 0 & $\tfrac{1}{2}b \Tgap$ \\
  1 & 1 & 0 & 0 & 1 &  $\tfrac{1}{2}b \Tgap$ \\
  1 & 1 & 0 & 1 & 0 &  $\tfrac{1}{2}d \Tgap$ \\
  1 & 1 & 0 & 1 & 1 & $d$ \\
  \hline
\end{tabular}
\end{center}
\smallskip

Hence the minimum probability that the left border $v_i$
and, by symmetry, the right border $v_j$ switch to cooperation is equal to
$\min \bigl\{\tfrac{1}{2}b,\tfrac{1}{2}d\bigr\} = \tfrac{1}{2}d.$ Hence we have
\begin{equation*}
\E[\sigma_t]\ \le \ \sigma_{t-1} - d\ .
\end{equation*}

This case is quite similar to Lemma~\ref{lem:absorptiontime},
and the result follows by a similar argument.
\end{proof}

\bigskip
\noindent \textbf{ Proof of Theorem ~\ref{thm:fastconvDEandFfinal} (Case - I): }
According to Lemma~\ref{lem:maxLengthDE}, the longest chain of d-runs separated
by singleton cooperators (non-barrier-segment) is
$O (\log n)$ w.h.p. Substituting this value for $\gamma$ in Lemma~\ref{lem:fastconvDandE} shows that the probability that
the all-cooperate is not reached in time $\tfrac{O(\log n)}{d(1- \varepsilon)} = O(\log n)$  is
at most  \[e^{-\tfrac{\varepsilon^2 O(\log n)}{2(1-\varepsilon)}} =  e^{-\tfrac{\varepsilon^2 c_1 \log n}{2(1-\varepsilon)}}= n^{-\tfrac{\varepsilon^2 c_1}{2(1-\varepsilon)}}\]  for any $\varepsilon > 0$ and a constant $c_1 > 0$. So, for suitable values of
$c_1$ and $\varepsilon$,  the above probability tends to zero as $n \to \infty$ .

Hence, if there is at least one barrier ($2_\textrm{c}$-run) at the beginning of the game,
the game converges to cooperation fast. Furthermore,
Lemma~\ref{lem:barrierpresence} shows that the initial configuration has a $2_\textrm{c}$-run
except for exponentially small failure probability, completing the proof.
\hfill $\Box$

\bigskip
\noindent\textbf{Region F ($T < S$)}
\medskip
\label{sec:F}

This region has been labelled F in Figure~\ref{fig:gamedomainsLabelled}.
The switching probabilities for this region are given in the table below.

\smallskip
\begin{center}
\begin{tabular}{|c|c|l|l|l|l|}
  \hline
 \multirow{2}{*}{$\CS_{i}$}  \Tgap & \multirow{2}{*}{$\CS_{i+1}$}& \multicolumn{4}{|c|}{$\CS_{i-1}\CS_{i+2}$} \\
\cline{3-6}
  &  & 00 & 01 & 10 & 11\\
\hline
    $0$ & $1$ & $+$  & $+$  & $+$ & $+$  \\
   \hline
    $1$ & $0$ & $0$ & $0$ & $0$ & $0$  \\
  \hline
\end{tabular}
\end{center}
\smallskip

It is clear from the table above that the evolution happening in this region is the opposite
to what happens in the region A (see Section~\ref{sec:A} for details).
More precisely, defectors are imitated by cooperators in region A, while cooperators are
imitated by defectors in region F. Hence, the cooperation evolves fast in this region and
the following lemma holds.

\begin{lem}
\label{lem:fastconvF}
Suppose $R_\textrm{d}$ is the longest d-run on the cycle when the game is started and
let $\ell(R_\textrm{d}) =\gamma$.
Provided $T < S$ , probability that all-cooperate state is not reached in
$\frac{\gamma}{c(1- \varepsilon)}$
generations is at most $e^{-\tfrac{\varepsilon^2\gamma}{2(1-\varepsilon)}}$ for any $\varepsilon >0$.
\end{lem}
\begin{proof}
Proof of this lemma is similar to the proof of Lemma~\ref{lem:fastconvA}. Let $\rho$ be
defined as the minimum  of the four possible switching probabilities. Then we get
\begin{equation*}
 \rho = \min \{a,b,c,d \} =c\  .
\end{equation*}
In this case, a d-run of length $\ell \,\,(1 \le \ell < n)$ is reduced in length
at both ends with probability at least $\tfrac{1}{2}\rho$ and a $1_\textrm{c}$-run is deleted with probability
at least $\rho$. Also, the absorption time here is the time it takes for the longest
d-run to be deleted. The rest of the proof is similar to that of Lemma~\ref{lem:fastconvA}.
\end{proof}

\bigskip
\noindent \textbf{
Proof of Theorem~\ref{thm:fastconvDEandFfinal} (Case II - $T < S$) : }
The result directly follows from Lemma~\ref{lem:fastconvF} and Lemma~\ref{lem:maxdrun}.
\hfill $\Box$

\bigskip
\noindent\textbf{A special case for B, C, D and E}

Consider the case where $1$'s and $0$'s appear in
the cycle at alternating locations. This configuration
will not be generated by the random initial configuration w.h.p. Because, as
proved in Lemma~\ref{lem:maxLengthAlternating}, the longest chain
of alternating 1's and 0's is $O(\log n)$ w.h.p.  However, it is worth
investigating this configuration as it yields some interesting outcome.

\begin{thm}
\label{lem:alternatingMinuses}
If the game ever reaches a state where every cooperator and every defector
on the cycle exist as singletons, the following statements hold with high
probability.

\begin{enumerate}
    \item the all-defect state is reached in time $O(1)$   with probability 1 in B and D.
 	 \item the all-cooperate state is reached, with probability strictly less than 1, in time $O(\log n)$ in C and in time $O(n)$  in E.
\end{enumerate}
 \end{thm}
\begin{proof}
$n$ has to be even for this scenario to occur. So, let $n = 2m$.
The configuration in question has alternating 1's and 0's throughout the cycle.
In this case, all the 0's appear as $1[0]10$
having switching probability of 0 while all the 1's appear as $0[1]01$ having
 switching probability of $\bar{c} >0$  (see Figure~\ref{fig:switchingprob2}). If all the 1's switch to 0  in the same generation, then the all-defect state
is reached in a single generation. However, the probability of that happening is $(\bar{c})^{m}$, which is small for large $m$ since $\bar{c} <1$ .

Now, consider the regions B and D. It is easily seen from their dynamics
on Figure~\ref{fig:dynamicsBCDE} that $1_\textrm{c}$-runs are always deleted. In a starting configuration of alternating 1's and 0's,
there are only $1_\textrm{c}$-runs present. Although, they all appear in the form $10[1]01$
in the initial configuration, the other forms $00[1]00$ and $00[1]01$ might be
generated in the subsequent generations. Singleton cooperators are then deleted in the expected time
of $1/\bar{a} = O(1)$ (see, Lemma~\ref{lem:shortRunsB} for the calculations involved).
This proves the first part of the theorem.

Now consider any 15 consecutive vertices in the initial configuration $v_{j-7} \ldots v_{j+7}$
such that $\CS_{j}=1$. The probability that all cooperators in
these vertices but $\CS_{j}$ switch to defection together is $(\bar{c})^6(1-\bar{c})$. Expected number
of such cases after the first generation is $n(\bar{c})^6(1-\bar{c})$. All these cases would have created
a $1_\textrm{c}$-run that has at least 7 defectors on both sides. Recall that $1_\textrm{c}$-run can
grow in C and E when they are bordered by at least two defectors (Lemma~\ref{lem:growingSingletonPlus}).
In the above case, once the $1_\textrm{c}$-runs grow to be a run of length 2
or 3, it still has at least 2 defectors at either side and can grow further (Lemma~\ref{lem:barrier}).
Thus, the required barrier for a guaranteed convergence to cooperation, i.e.\ a $4_\textrm{c}$-run
for C and a $2_\textrm{c}$-run for E, will be created in time $O(1)$. Then the results follow
from Theorem~\ref{thm:fastconvBandCfinal} and Theorem~\ref{thm:fastconvDEandFfinal}.
It was remarked above that the game might converge to the all-defect  state even in C and E
with probability $(\bar{c})^{m}$.
So, the probability that all-cooperate is reached is less or equal to $1-(\bar{c})^{m} < 1$.
\end{proof}

\subsubsection{Borders}
\medskip
\label{sec:Borders}

We now  analyse the behaviour of the games which lie on
borders between regions.
On these borders, one might expect to see a mixed result of
the two regions that the border separates. The results below
show this intuition is wrong.

\bigskip
\noindent \textbf{The Line $T = S+1$}
\medskip
\label{sec:BorderAB}

This is the border between regions A and B. Let
$\mathcal{B}_{AB}$ denote this border. The
table below shows which of the switching probabilities  are
zero for $\mathcal{B}_{AB}$.

\smallskip
\begin{center}
\begin{tabular}{|c|c|l|l|l|l|}
  \hline
 \multirow{2}{*}{$\CS_{i}$}  \Tgap & \multirow{2}{*}{$\CS_{i+1}$}& \multicolumn{4}{|c|}{$\CS_{i-1}\CS_{i+2}$} \\
\cline{3-6}
  &  & 00 & 01 & 10 & 11\\
\hline
    $0$ & $1$ & $0$  & $0$  & $0$ & $0$  \\
   \hline
    $1$ & $0$ & $+$ & $+$ & $0$ & $+$  \\
  \hline
\end{tabular}
\end{center}
\smallskip

The following observations that can be verified using the table above will help
our analysis.

\begin{itemize}
    \item A $1_\textrm{c}$-run is always deleted.
    \item An $\ell_\textrm{c}$-run  $(\ell \ge 2)$ can never grow.
    \item A $1_\textrm{d}$-run can become a $2_\textrm{d}$ or a $3_\textrm{d}$-run.
    \item An $\ell_\textrm{d}$-run  $(\ell \ge 2)$ cannot grow if it is  bordered
          by at least two cooperators.

\end{itemize}

In essence, the only changes that happen in any generation are: a $1_\textrm{c}$-run is deleted,
and  a $1_\textrm{d}$-run becomes a $2_\textrm{d}$-run or a $3_\textrm{d}$-run.
This suggest that cooperators and defectors can coexist in a steady
state as long as they are not singletons.

\begin{thm}\label{thm:borderAB}
If~$T=S+1$, the game reaches steady state in time $O(\log n)$
with high probability.
The steady state contains d-runs and c-runs of length at least $2$,
and the proportion of cooperators in the steady state
is less or equal to the initial proportion.
\end{thm}
\begin{proof}
In the evolution phase, what happens is the elimination of singleton
cooperators and extension of singleton defectors. We note the followings when
we look at c-runs of different lengths:
\begin{itemize}
    \item An $\ell_\textrm{c}$-run ($\ell \ge 2$) having at least two defectors
          on both sides remain unchanged.
    \item An $\ell_\textrm{c}$-run ($\ell \ge 3$) having at least two defectors
          on one side and a $1_\textrm{d}$-run on the other side can be reduced
          in length  by at most 1 and become stable as in the previous case. The probability
          that a cooperator is deleted in a generation is $p_1=\tfrac{1}{2}\bar{d}$.
    \item An $\ell_\textrm{c}$-run ($\ell \ge 4$) having $1_\textrm{d}$-runs at either sides can be reduced
          in length   by at most 2 and become stable as in the first case. The probability
          that a cooperator is deleted in a generation is at least $p_2=\tfrac{1}{2}\bar{d}$.
	  \item All other cases of c-runs are potentially deleted in the worst case. They are:
				\begin{itemize}
    				\item  $1_\textrm{c}$-runs. In this case, the cooperator is deleted with
 probability at least $p_3= \min\bigl\{\bar{a}, \bar{c}, \\ \tfrac{1}{2}(\bar{a}+\bar{c}) \bigr\} = \bar{a}.$
					\item $2_\textrm{c}$-runs having at least one of its neighbour as $1_\textrm{d}$-run. In this case,
					a cooperator is deleted with  probability at least
					$p_4 = \min \bigl\{ \tfrac{1}{2}\bar{d}, \bar{d}-\tfrac{1}{4}{\bar{d}}^2 \bigr\} = \tfrac{1}{2}\bar{d}$.
					\item $3_\textrm{c}$-runs having both neighbours as $1_\textrm{d}$-runs. In this case, a cooperator is deleted
					with probability at least $p_5 = \bar{d}-\tfrac{1}{4}{\bar{d}}^2$.
				\end{itemize}
\end{itemize}
Let $\rho_{\textrm{ab}} =\min \{  p_1, p_2, p_3,p_4,p_5 \}$. Hence,
all cooperators are deleted with probability at least $\rho_{\textrm{ab}}$ in any generation. Maximum number of
cooperators that any c-run can loose is 3, thus the absorption time is the time it
takes for a $3_\textrm{c}$-run to be deleted at the slowest rate.
Then, by Lemma~\ref{lem:absorptiontime}, for any $\varepsilon >0$, the probability that a steady
state is not reached in time $\tfrac{3}{\rho_{\textrm{ab}}(1-\varepsilon)} = O(1)$ is
a constant.
 Hence, after time $O(\log n)$, the probability that a steady state is not reached is at most $n^{-k}$, for any $k$.
\end{proof}

\bigskip
\noindent\textbf{The Line $T = \tfrac{S+1}{2}$}
\medskip
\label{sec:BorderBC}

The line $T = \tfrac{S+1}{2}$ separates the regions B $\cup$ C and D $\cup$ E.
Let us call this $\mathcal{B}_{BD}$. The corresponding switching probabilities
are given in the table below.

\smallskip
\begin{center}
\begin{tabular}{|c|c|l|l|l|l|}
  \hline
 \multirow{2}{*}{$\CS_{i}$}  \Tgap & \multirow{2}{*}{$\CS_{i+1}$}& \multicolumn{4}{|c|}{$\CS_{i-1}\CS_{i+2}$} \\
\cline{3-6}
  &  & 00 & 01 & 10 & 11\\
\hline
    $0$ & $1$ & $0$  & $+$  & $0$ & $0$  \\
   \hline
    $1$ & $0$ & $+$ & $+$ & $0$ & $0$  \\
  \hline
\end{tabular}
\end{center}
\smallskip

It is readily verified from the above table that, on $\mathcal{B}_{BD}$, the
only way the defectors can spread is through deleting singleton cooperators. At the same time,
no singleton cooperators are created during the evolution. Hence, deleting singleton cooperators is
the only setback in the process of emergence of cooperation. In order to calculate the worst case
absorption time, we divide the process into two phases.

\begin{description}
    \item[Phase I] The singleton cooperators are allowed to disappear first, suppressing all 	
other favourable developments.
    \item[Phase II] The rest of the evolution, assuming that
	there are no singleton cooperators. In this phase, an $\ell_\textrm{d}$-run
($\ell \ge 2$) reduces in length until it is of length $0$ or $1$.
Singleton defectors cannot be deleted on $\mathcal{B}_{BD}$.
\end{description}

Note that, in the actual process though, both Phase I and II happen simultaneously.
So the above gives an upper bound on the absorption time. First,
we calculate the time required for Phase I.

\begin{lem}
\label{lem:singletonsRemD}
All $1_\textrm{c}$-runs are removed in  time $O(1)$.
\end{lem}
\begin{proof}
A $1_\textrm{c}$-run or singleton can exist in the cycle in three different
surroundings, having a different probability of removal accordingly:

\begin{itemize}
    \item  The $1_\textrm{c}$-run in $00[1]00$ is deleted with probability $\bar{a}$.
     \item The $1_\textrm{c}$-run in $00[1]01$ is deleted with probability $\tfrac{1}{2}(\bar{a}+\bar{c})$.
\item The $1_\textrm{c}$-run in $10[1]01$ is deleted with probability $\bar{c}$.
\end{itemize}

As the update rule is synchronous, the worst case time to remove
all singletons is determined by the smallest probability of removal. That
is,
\begin{equation*}
\varphi_1 = \min \bigl\{\bar{a}, \tfrac{1}{2}(\bar{a}+\bar{c}), \bar{c} \bigr\} = \bar{a},
\end{equation*}
as we have $\bar{a} \le \bar{c}$. Using the geometric distribution, the worst expected
time required to remove all singletons is
\begin{equation*}
\E[T_s] = \dfrac{1}{\varphi_1} = \dfrac{1}{\bar{a}} = O(1)\ .
\end{equation*}
\end{proof}

Next, we analyse Phase II. Here we analyse d-runs. As there
are no $1_c$-runs present on the cycle in this phase, the length of a d-run
will decrease until it is removed completely or it becomes
a singleton. Since the updates are simultaneous, the time
taken to reduce all d-runs is determined by the longest d-run
present in the cycle.

\begin{lem}
\label{lem:absorptionTImeC}
Suppose $R_\textrm{d}$ is the longest d-run on the cycle when Phase II is
started and let $\ell(R_\textrm{c}) =\gamma$.
If $T = \tfrac{S+1}{2}$ then, for any $\varepsilon >0$, the probability that a steady state
is not reached in
$\frac{\gamma }{b(1- \varepsilon)}$
generations is at most $e^{-\tfrac{\varepsilon^2\gamma}{2(1-\varepsilon)}}$. In the steady state, cooperators and defectors
coexist. The defector runs must be singletons and the cooperator runs cannot be singletons.
\end{lem}
\begin{proof}

The key idea of this proof is, all d-runs are bordered by a c-run of length at least
2 in Phase II. Consequently, the vertices at either end of a
d-run switch to cooperation with probability $\tfrac{1}{2}b$. Hence the expected
decrease in length of the d-runs is equal to $b$.

Also note that singleton defectors cannot be removed. Hence, all
longer d-runs are either turned into a $1_\textrm{d}$-run or are removed.
The longest d-run $R_\textrm{d}$ requires the longest time to delete or reduce to a singleton.
When $R_\textrm{d}$ is deleted, or made into a $1_\textrm{d}$-run, the steady state will be reached. Let $\ell_t$ be the
length of $R_\textrm{d}$ after t steps. Then we get
\begin{equation*}
\E[\ell_t]\ \le\ \ell_{t-1} - b\ =\ \gamma -bt\ .
\end{equation*}

Now, the result directly follows from Lemma~\ref{lem:absorptiontime}.
\end{proof}

\begin{thm}
\label{thm:fastconvBorderBD}
If  $T = \tfrac{S+1}{2}$, a steady state
is reached in time $O(\log n)$  w.h.p. In the steady state, cooperators and defectors
coexist. The defector runs must be singletons and the cooperators runs cannot be singletons.
\end{thm}

\begin{proof}
The worst case absorption time is
obtained by summing the time required for Phase I and Phase II. Lemma~\ref{lem:singletonsRemD}
shows the worst case time for Phase I is $O(1)$. Lemma~\ref{lem:absorptionTImeC} proves that,
if $T = \tfrac{S+1}{2}$, the probability that a steady state is not reached in
$\frac{\gamma}{b(1 - \varepsilon)}$
generations is at most $e^{-\tfrac{\varepsilon^2\gamma}{2(1-\varepsilon)}}$ for any $\varepsilon >0$, where $b = \tfrac{S+1-T}{2\alpha}$ and
$\gamma$ is the size of the longest d-run at the beginning of Phase II.

So we merely need to bound the length $\gamma$ of the longest d-run
at the beginning of Phase II. We observe that the removal of singleton cooperators in Phase I will join d-runs together.
However, we showed in Lemma~\ref{lem:maxLengthDE} that the longest chain of defectors interleaved
with singleton cooperators is still $O(\log n)$ w.h.p.
Combining this with the estimate $\frac{\gamma}{b(1 - \varepsilon)}$ completes the proof.
\end{proof}

\bigskip
\noindent\textbf{The Line $T = S$}
\medskip
\label{sec:BorderEF}

The line $T = S$ on the $ST$-plane, denoted by $\mathcal{B}_{FE}$, separates the regions F and E. The table below
shows which of the switching probabilities are zero in this case.

\smallskip
\begin{center}
\begin{tabular}{|c|c|l|l|l|l|}
  \hline
 \multirow{2}{*}{$\CS_{i}$}  \Tgap & \multirow{2}{*}{$\CS_{i+1}$}& \multicolumn{4}{|c|}{$\CS_{i-1}\CS_{i+2}$} \\
\cline{3-6}
  &  & 00 & 01 & 10 & 11\\
\hline
    $0$ & $1$ & $+$  & $+$  & $0$ & $+$  \\
   \hline
    $1$ & $0$ & $0$ & $0$ & $0$ & $0$  \\
  \hline
\end{tabular}
\end{center}
\smallskip

Here, no cooperator ever becomes a defector. On $\mathcal{B}_{FE}$, therefore, an $\ell_\textrm{c}$-run $(\ell \ge 1)$ is a barrier. Yet, a $1_\textrm{c}$-run adjacent to a $1_\textrm{d}$-run on either side
cannot grow and help the convergence to cooperation. Therefore, a barrier
for $\mathcal{B}_{FE}$ is an  $\ell_\textrm{c}$-run where $\ell \ge 2$ or a $1_\textrm{c}$-run adjacent
to two defectors. A non-barrier-segment is bounded by two such barriers.
The non-barrier segments are eliminated by successively deleting the end vertices.

There are clearly two types of non-barrier-segment for  $\mathcal{B}_{FE}$:
a d-run, and a run of alternating 1's and 0's. In both cases, the all-cooperate
state is attained by the expansion of the barriers.

\begin{lem}
\label{lem:fastconvBorderFE}
Suppose $R_{cd}$ is the longest non-barrier-segment and let $\gamma$
be its initial length. Then,  if $T = S$, the probability that the all-cooperate state is not
reached in
$\frac{\gamma}{d(1- \varepsilon)}$
generations is at most $e^{-\tfrac{\varepsilon^2\gamma}{2(1-\varepsilon)}}$,
for any $\varepsilon > 0$.
\end{lem}

\begin{proof}
The proof is very similar to that of Lemma~\ref{lem:fastconvDandE}, so is omitted.
\end{proof}

What remains is to determine the length of the longest non-barrier-segment.
Obviously, in the worst case, at the beginning of the game there will be only one barrier and the
 non-barrier-segment will be of length $n-1$ .
Then, by Lemma~\ref{lem:fastconvBorderFE},
the worst-case absorption time is $O(n)$ with high
probability. But,  when the game is started with a random configuration, the longest non-barrier-segment is only $O(\log n)$.

\begin{thm}
\label{thm:fastconvBorderFE}
If $T=S$, the all-cooperate
state is reached in time $O(\log n)$,  with high probability.
\end{thm}

\begin{proof}
The longest d-run present at the beginning of the game is $O(\log n)$ w.h.p by Lemma~\ref{lem:maxcrun}, and
the longest chain of alternating defectors and cooperators is $ O( \log n)$ w.h.p
by Lemma~\ref{lem:maxLengthAlternating}.
Hence, the longest non-barrier-segment is $O(\log n)$ w.h.p.
Substituting $\gamma$ in Lemma~\ref{lem:fastconvBorderFE} completes the proof.
\end{proof}

\begin{rem}
It is readily verified that alternating 1's and 0's  throughout the cycle
is a steady state for  $\mathcal{B}_{FE}$. But, this state can never be reached
unless it is the initial configuration.
\end{rem}

\bigskip
\noindent\textbf{The Line $T = 2S$}
\medskip
\label{sec:BorderED}

$\mathcal{B}_{ED}$  is the line between the regions E and D.
The switching probability table given below shows that the dynamics on this line are close to the dynamics of E. In fact, the only difference
is that singleton cooperators, having at least two defectors adjacent, can grow in E, but not on $\mathcal{B}_{ED}$. Even so, we analysed the evolution of cooperation for E without considering that singletons can contribute to the evolution of cooperation. Only the growth of barriers was considered. Hence, the proof of Theorem~\ref{thm:fastconvDEandFfinal} also applies here.

\smallskip
\begin{center}
\begin{tabular}{|c|c|l|l|l|l|}
  \hline
 \multirow{2}{*}{$\CS_{i}$}  \Tgap & \multirow{2}{*}{$\CS_{i+1}$}& \multicolumn{4}{|c|}{$\CS_{i-1}\CS_{i+2}$} \\
\cline{3-6}
  &  & 00 & 01 & 10 & 11\\
\hline
    $0$ & $1$ & $0$  & $+$  & $0$ & $+$  \\
   \hline
    $1$ & $0$ & $0$ & $+$ & $0$ & $0$  \\
  \hline
\end{tabular}
\end{center}
\smallskip

\begin{thm}
\label{thm:fastconvBorderED}
On the line $T= 2S$, the all-cooperate state will be reached in time $O(\log n)$,  with high probability. \qed
\end{thm}

\subsection{Summary}
\label{sec:summary}

We have proved rigorously that the games converge fast for all $-1 \le S \le 1$ and  $0 \le T \le 2$.
We have done this by grouping the games based on various relations between the payoffs. The results are summarised below.
\begin{description}
    \item [1.~$0 \le T < \tfrac{S+1}{2}$ and $T \ne 2S$:] This encompasses regions D, E, F and the
	border between E and F. In these  regions, the all-cooperate state is reached in time $O(\log n)$.
    \item [2.~$\tfrac{S+1}{2} < T < S+1$:] This encompasses regions B and C. In these  regions,
	the all-cooperate state is reached in time $O(n)$.
    \item [3.~$S+1 < T \le 2$:] This contains region A. In this  region, the all-defect state is
	reached in time $O(\log n)$.
    \item [4.~$T = 2S$:] This is the border between B\,$\cup$\,D and C\,$\cup$\,E. On this line, the
	all-cooperate state is reached in time $O(n)$.
    \item [5.~$T = \tfrac{S+1}{2}$:] This is the border between B\,$\cup$\,C and D\,$\cup$\,E. On this line,
	a steady state is reached in time $O(\log n)$. Here cooperators and defectors can coexist indefinitely.
    \item [6.~$T = S+1$:] This is the border between A and B. On this line, a steady state is
	reached in time $O(\log n)$. Here cooperators and defectors can coexist indefinitely.
\end{description}

The coexistence of cooperators and defectors when $T= S+1$ can be explained
as follows. When a defector earning $T$ is adjacent to a cooperator earning
$S+1$, the switching probability for both players is zero (e.g.\ $11110000$, where
this applies to the middle cooperator and defector).
There is a player following the same strategy on one side and a player following a
different strategy, but with the same payoff, on the other side. Hence, there is
no incentive for imitation. The coexistence on $2T = S+1$ can be explained
in the same way. This happens when a singleton defector is between
two runs of cooperators (e.g.\ $1110111$).

\begin{rem}
 As mentioned earlier, another version of the imitation strategy on the cycle was studied
analytically in~\cite{Ohtsuki2006}, by calculating fixation probabilities.
The fixation probabilities can be calculated easily using our arguments.
Let $P_\textrm{c}$ be the fixation probability for the cooperator, i.e.\  the probability that
the game converges to cooperation when a single cooperator is added to $n-1$ defectors on the cycle.
Then we have $P_\textrm{c} = 0$ for regions A, B and D; and $P_\textrm{c} = 1$ for regions C, E and F. However, the games converge to cooperation
in all but region A when a maximum of four cooperators are introduced. This illustrates the drawbacks
of analysis based on fixation probabilities.
\end{rem}

\section{Imitation on the complete graph}
\label{sec:completeGraph}

In this section, we analyse the imitation update rule on the complete graph
$K_n$. Each vertex of the graph plays the game with all other vertices.
Hence, all cooperators receive the same accrued payoff in a given generation,
and have the same switching probability. The same applies to the defectors.

Now, let $\delta_t$ denote the number of cooperators in $K_n$ at time $t$.
Let $\Py^t_\textrm{d}$ and $\Py^t_\textrm{c}$ denote the total payoff obtained
 by a defector and a cooperator respectively, at time $t$.
As before, using the normalisation $R=1$ and $P=0$, we get
\begin{equation*}
   \Py^t_x = \begin{cases}
        \delta_tT & \text{if $x = \textrm{d}$},\\
        (\delta_t-1) + (n-\delta_t)S & \text{if $x = \textrm{c}$}.
        \end{cases}
\end{equation*}

Let $\Delta_{yx} = \Py^t_y - \Py^t_x$ where $x,y \in \{\textrm{c},\textrm{d}\}$. Figure~\ref{fig:switchingprobKn} gives the values
of $\Delta_{yx}$ for different values of $x$ and $y$. Then
the switching probability for a vertex adopting the strategy $x$ at time $t$, which
has chosen a vertex adopting $y$ for imitation, is
\[p_x^t = \max \biggl\{0, \dfrac{\Py_y^t - \Py_x^t}{(n-1)\alpha} \biggr\} =
 \max \biggl\{0, \dfrac{\Delta_{yx}}{(n-1)\alpha}  \biggr\} ,  \]
where $\alpha =  \max\{T, 1\}- \min\{S,0\}$. Here the denominator
ensures that $p_i \in [0,1].$ Note that only $p_\textrm{c}^t$ or $p_\textrm{d}^t$ can be positive at any $t$.

\begin{figure}[ht]
\centering
\begin{tabular}{|c|c|c|}
  \hline
  $x$ & $y$ & $\Delta_{yx}$\\
  \hline
   d& d & 0\\
    d & c &   $\delta_t (1-S-T) + nS-1$ \\
    c & d &  $-\delta_t (1-S-T) - nS+1$ \\
    c & c & 0 \\
  \hline
\end{tabular}
\caption[Payoff difference between two players on a complete graph]{Payoff difference between two players
adopting $x$ and $y$ at time $t$ on a complete graph. Here, d stands for  defection and c stands for a cooperation.}
\label{fig:switchingprobKn} 
\end{figure}

\begin{thm}
\label{thm:allCoopKn}
Let $\theta_t$ denote the fraction of cooperators on the complete
graph $K_n$ at time $t$. Suppose, for large $n$, that $\theta_0$, $S$ and $T$  satisfy
\begin{equation*}
    T\ <\ \min\left\{1,\,1+\frac{(1-\theta_0)}{\theta_0 }\,S\right\}.
\end{equation*}
and that there is at
least one cooperator at the beginning of the game (i.e.\ $\theta_0 \ge \tfrac{1}{n}$). Then,
for any $\varepsilon >0$, the probability that the all-cooperate state is not reached in time
\[\dfrac{1}{\varphi_\textrm{d}\theta_0 } \log \left( \dfrac{ n(1-\theta_0)}{\varepsilon} \right)\]
is at most $\varepsilon$, where $\varphi_\textrm{d}$ satisfies
\[\varphi_\textrm{d}\  <\  \min\left\{\dfrac{\theta_0(1-S-T) + S}{\alpha}, \dfrac{1-T}{\alpha}\right\}, \quad \mbox{with} \quad \alpha =  1- \min\{S,0\}\ .\]

\end{thm}

\begin{proof}
Let $\delta_t = n\theta_t$.
A defector switches
to cooperation in any generation with some positive probability $p_\textrm{d}^t$ only a
cooperator with higher payoff is chosen for imitation. That is, a defector
switches to cooperation at time $t$ only if $\Delta_{\textrm{cd}} > 0$. Hence,
from Figure~\ref{fig:switchingprobKn}, the required condition is
\[\Delta_{\textrm{cd}}= \delta_t (1-S-T) + nS-1 > 0\ .\]

Substituting  $n\theta_t$ for $\delta_t$ and rearranging,  we get
\begin{equation*}
 T\ <\ 1+  S \biggl(\dfrac{1}{\theta_t} -1\biggr) - \dfrac{1}{n\theta_t}
   \ =\  1+  \dfrac{1}{\theta_t} \biggl(S- \dfrac{1}{n}\biggr) - S,
\end{equation*}
so, for large enough $n$, this condition is
\begin{equation}
\label{eq:allcoopcon1}
 T\ <\ 1+  \dfrac{(1-\theta_t)}{\theta_t} S\ .
\end{equation}
As defectors become cooperators, $\theta_t$ increases. If $S \leq 0$,
inequality~\eqref{eq:allcoopcon1} holds as $\theta_t$ increases up to $1$.
But, if $S > 0$, the inequality holds only until $\theta_t=1$ provided $n$ is large enough and
\begin{equation}
\label{eq:allcoopcon2}
 T < 1\ .
\end{equation}
Thus, if $\theta_0$, $S$ and $T$ satisfy inequalities~\eqref{eq:allcoopcon1} and
\eqref{eq:allcoopcon2} at $t=0$, then $\theta_t$ increases from
$\theta_0$ to 1.  Let $\varphi_\textrm{d} > 0$ be a lower bound on the switching probability at any step during this process. That is,
\[ \varphi_\textrm{d}\  <\ \inf_{ \theta_0 \le \theta_t < 1} p_\textrm{d}^t \ =\   \dfrac{\theta_t(1-S-T) + S}{\alpha}= \min\left\{\dfrac{\theta_0(1-S-T) + S}{\alpha}, \dfrac{(1-S-T) + S}{\alpha}\right\},\]
where $\alpha = \max\{ T, 1\} - \min \{S,0 \} = 1 - \min \{S,0 \}$.
Let $\gamma_t$ denote the number of defectors at time $t$. Then, we can
determine the expected value of $\gamma_{t}$ in terms of $\gamma_{t-1}$.
\begin{equation*}
\E[\gamma_{t} \mid \gamma_{t-1}] \le \gamma_{t-1} - \gamma_{t-1}\Bigl(\dfrac{n-\gamma_{t-1}}{n-1} \Bigr) \varphi_\textrm{d} < \gamma_{t-1} - \gamma_{t-1}\Bigl(\dfrac{n-\gamma_{t-1}}{n} \Bigr) \varphi_\textrm{d} = \gamma_{t-1} \Bigl(  1 - \varphi_\textrm{d}\theta_{t-1} \Bigr)\ .
\end{equation*}

We know that, under conditions \eqref{eq:allcoopcon1} and
\eqref{eq:allcoopcon2}, $\theta_{t} \ge \theta_0$ for any $t \ge 1$.
Substituting this in the above inequality, we get
\begin{equation*}
\E[\gamma_{t} \mid \gamma_{t-1}] \le \gamma_{t-1} (  1 - \varphi_\textrm{d}\theta_0) \\
\end{equation*}

Thus, by the rule of total expectation, we have
\begin{equation*}
\E[\gamma_{t}] \le  (  1 - \varphi_\textrm{d} \theta_0) \E[\gamma_{t-1}]\ .
\end{equation*}

Applying this iteratively for $1$ to $t-1$, we obtain
\begin{equation*}
\E[\gamma_t] \le (  1 - \varphi_\textrm{d}\theta_0)^t \E[\gamma_0] = (  1 - \varphi_\textrm{d} \theta_0)^t \gamma_0\ .
\end{equation*}

So, for any $\varepsilon >0$, when
\begin{equation}
\label{e:timefastc}
t >  \dfrac{1}{\varphi_\textrm{d}\theta_0 } \log \left( \dfrac{\gamma_0}{\varepsilon} \right),
\end{equation}
we have $\E[\gamma_t] \le \varepsilon$. We know that any nonzero value of $\gamma_t$ is at least $1$.
Using Markov's inequality, we obtain
\begin{equation*}
\Pr[\gamma_t \ne 0]\ =\ \Pr[\gamma_t \ge 1]\ \le\ \E[\gamma_t]\ \le\ \varepsilon\ . \end{equation*}

Finally, substituting $\gamma_0 = n(1-\theta_0)$ into \eqref{e:timefastc}  completes the proof.
\end{proof}

\begin{thm}
\label{thm:allDefectKn}

Let $\theta_t$ denote the fraction of cooperators on the complete
graph $K_n$ at time $t$. Suppose, for large $n$, that $\theta_0$,
$S$ and $T$  satisfy
\begin{equation*}
    T\ >\ 1+  \dfrac{1-\theta_0}{\theta_0} S,\ \ \mathrm{and}\ \
    S \le 0,
\end{equation*}
and that there is at least one defector at the beginning of
the game (i.e.\ $(1-\theta_0) \ge \tfrac{1}{n}$). Then,  for any $\varepsilon$, the probability
that the all-defect state is not reached in time
\[\dfrac{1}{\varphi_\textrm{c}(1-\theta_0) } \log \left( \dfrac{ n\theta_0}{\varepsilon} \right)\]
is at most $\varepsilon$, where $\varphi_\textrm{c}$ satisfies
\[\varphi_\textrm{c}  <  \min\left\{ \dfrac{T -(n - 1)S}{(n-1)\alpha}, \dfrac{\theta_0(T+S-1) - S}{\alpha} \right\} , \quad \mbox{with} \quad \alpha =  \max\{T, 1\}- S\ .\]
\end{thm}

\begin{proof}
This can be proved in the same way as Theorem~\ref{thm:allCoopKn}. Hence,
only the main differences are highlighted here.

Let $\delta_t = n\theta_t$ as earlier. For a cooperator to switch to defection at time $t$, the switching probability $p^t_\textrm{c}$ has
to be positive. This is  possible only if $\Delta_{\textrm{dc}}$ is positive. Hence,
from Figure~\ref{fig:switchingprobKn}, we have
\begin{equation*}
\Delta_{\textrm{dc}}= -\delta_t (1-S-T) - nS+1 > 0\ .
\end{equation*}

Substituting $\delta_t = n\theta_t$ into the above inequality and solving
for $T$ yields
\begin{equation*}
 T\ >\ 1+  S \biggl(\dfrac{1}{\theta_t} -1\biggr) - \dfrac{1}{n\theta_t}
 \  =\  1+  \dfrac{1}{\theta_t} \biggl(S- \dfrac{1}{n}\biggr) - S\ .
\end{equation*}
For large enough $n$, this is
\begin{equation}
\label{eq:alldefectcon1}
 T\ >\ 1+  \dfrac{1-\theta_t}{\theta_t}S\ .
\end{equation}
As cooperators become defectors, $\theta_t$ decreases. Inequality~\eqref{eq:alldefectcon1}
continues to hold during this process if
\begin{equation}
\label{eq:alldefectcon2}
S \le 0\ .
\end{equation}
So, if inequalities \eqref{eq:alldefectcon1} and
\eqref{eq:alldefectcon2} hold initially,
the game converges to defection.
The proof can be completed by estimating $\E[\delta_t]$ in terms of $\delta_0$, as in Theorem~\ref{thm:allCoopKn}.
\end{proof}

\begin{cor}
\label{cor:steadystateline}
Let $T, S$, and $\theta_0$ be such that the cooperators
and defectors receive equal payoffs at the start of the game. That is,
\begin{equation*}
T =  1+  \dfrac{1}{\theta_0} \biggl(S- \dfrac{1}{n}\biggr) - S,\ \
\mbox{or equivalently}\ \ \delta_0 = \dfrac{nS-1}{T+S-1}\ .
\end{equation*}
Then all players maintain their initial strategy indefinitely.

This is the line in the ST plane which divides the regions where the all-cooperate state is reached fast (see Theorem~\ref{thm:allCoopKn}) and the all-defect state is reached fast (see Theorem~\ref{thm:allDefectKn}).
\end{cor}

Next we deal with the remaining region. This corresponds roughly
to the Snowdrift game.

\begin{thm}
\label{thm:coexisitenceKn}
Let $\delta_t$ denote the number of cooperators at time $t$. Then, the following statements fail with probability exponentially small in $n$.
If $\delta_0$ is outside the
range $\delta^{*} \pm n^{1/2+\varepsilon}$, for any constant $\varepsilon >0$, $S > 0$ and $T \ge 1$, then $\delta_t$ reaches this range in time $O(\log n)$  where
\begin{equation*}
\delta^{*}\ =\ \dfrac{nS - 1}{(S+T-1)}\ .
\end{equation*}
Thereafter, if $\delta^{*}$ is an integer, $\delta_t$ becomes
$\delta^{*}$ in time $O(\log n \log \log n)$  and remains with $\delta_t = \delta^{*}$  forever. If  $\delta^{*}$ is not an integer, then either the all-cooperate
or the all-defect state will eventually be reached, after $n^{O(n)}$ time.
\end{thm}

\begin{proof}
If  $\delta_t (1-S-T) + nS-1 > 0$, defectors in $K_n$ become cooperators with non-null probability  while the cooperators remain as cooperators. Note that we have $S+ T > 1$. Hence the above inequality can be rewritten as
\begin{equation*}
\delta_t < \dfrac{nS-1}{S+T-1} =\delta^{*}\ .
\end{equation*}

Similarly, if the following inequality is satisfied, the
cooperators become defectors with non-null probability while
the defectors remain as defectors.
\begin{equation*}
\delta_t > \dfrac{nS-1}{S+T-1} = \delta^{*}\ .
\end{equation*}

Finally, it can easily be verified that if $\delta_t = \delta^{*}$, the
switching probabilities of both cooperators and defectors are zero.
In short, $\delta_t$ oscillates around $\delta^{*}$  for a given $S$ and $T$ as follows:
\begin{itemize}
     \item if $\delta_t < \delta^{*}$, then $\delta_{t+1} \ge \delta_t$
    \item if $\delta_t > \delta^{*}$, then $\delta_{t+1} \le \delta_t$
     \item if $\delta_t = \delta^*$, then, $\delta_{t+1} = \delta^{*}$.
	 Note that this can only happen if $\delta^{*}$ is an integer.
\end{itemize}

First, suppose $\delta_t > \delta^{*}$. Then, we have that $\delta_t - \delta_{t+1}$ is binomial Bin$(\delta_t,p_t)$, where
\begin{equation}
\label{eq:prob1}
 p_t\ =\  \dfrac{n-\delta_t}{n-1}\, \dfrac{\delta_t(S+T-1) -nS+1}{T(n-1)}\ .
\end{equation}
So we have
\begin{equation}\label{eq:drift1}
\E[\delta_{t+1}]\ =\ \delta_t  -  \delta_t\dfrac{n-\delta_t}{n-1}\,\dfrac{\delta_t(S+T-1) -nS+1}{T(n-1)}\ .
\end{equation}
Similarly, for the case where $\delta_t < \delta^{*}$, we have $\delta_{t+1} - \delta_{t}$ is Bin$(n-\delta_t,\rho_t)$, where
\begin{equation}
\label{eq:prob21}
 \rho_t\ =\  \dfrac{\delta_t}{n-1}\, \dfrac{-(\delta_t(S+T-1) -nS+1)}{T(n-1)}\ ,
\end{equation}
and
\begin{equation}\label{eq:drift21}
\E[\delta_{t+1}]\ =\ \delta_t  -  (n-\delta_t)\dfrac{\delta_t}{n-1}\,\dfrac{\delta_t(S+T-1) -nS+1}{T(n-1)}\ .
\end{equation}
It is clear from \eqref{eq:drift1} and \eqref{eq:drift21} that the drift
towards $\delta^{*}$ is symmetrical. The rest of the proof, therefore,
considers only the case where $\delta_t > \delta^{*}$. Now,
let $\Delta_t$ be the number of cooperators with binomial
distribution Bin$(\Delta_t, \gamma_t)$, where
\begin{equation}
\label{eq:prob31}
\gamma_t\ =\ \dfrac{\Delta_t(S+T-1) -nS+1}{T(n-1)}\ .
\end{equation}
Let $I$ denote the interval $\delta^* \pm n^{1/2+\varepsilon}$ for any constant
$\varepsilon$. Suppose $\Delta_\tau = \delta_\tau > \delta^* + n^{1/2+\varepsilon}$. Let $T = \min \{ t > \tau \mid \delta_\tau \in I \}$. Then, the process $\delta_\tau \ldots \delta_T$ is stochastically dominated by $\Delta_\tau \ldots \Delta_T$
since we have $\gamma_t > p_t$. We now prove that the probability
that $\delta_t$ becomes less than $\delta^{*}$ is exponentially
small when $t \le T$.
Let $\beta_t=\Delta_t/(n-1)$, and
\[ \beta^*\ =\ \frac{(nS-1)}{(n-1)(S+T-1)}\ \sim\ \frac{S}{S+T-1},\]
so \eqref{eq:prob31} becomes
\begin{equation}
\label{eq:prob32}
 \gamma_t\ =\  (\beta_t -\beta^*)\frac{(S+T-1)}{T}\ =\ (\beta_t-\beta^*)s,
\end{equation}
where \[s = \frac{(S+T-1)}{T},\] so $s$ is bounded away from $0$ and $1$.
Also we have
\begin{equation}
\label{eq:drift31}
\E[\beta_{t+1}] = \beta_t  - \beta_t (\beta_t-\beta^*)s\,.
\end{equation}
Let $\Delta^* = (n-1)\beta^*$. While $\Delta_t\geq \Delta^*$, we have $\E[\Delta_{t+1}] \ge \Delta^*$ since
\[  \E[\Delta_{t+1}] = \Delta_t - \Delta_t  (\beta_t-\beta^*)s > \Delta^* \]
holds whenever $\beta_t s < 1$ which is always true.
Let $(1-x)\E[\Delta_{t+1}] = \Delta^*.$
Then, from~\eqref{eq:drift31} and using $\beta^* \le \beta_t \le 1$, we obtain
\begin{align}
x &= 1 - \dfrac{\Delta^*}{\Delta_t -\Delta_t  (\beta_t-\beta^*)s}  \nonumber\\
&= 1 - \dfrac{\beta^*}{\beta_t-\beta_t  (\beta_t-\beta^*)s} \nonumber \\
&= \dfrac{\beta_t -\beta^*}{\beta_t} \dfrac{1-\beta_t s}{1-  \beta_ts+\beta^*s}  \nonumber\\
&\ge (1- s)(\beta_t - \beta^*) = \dfrac{(1-S)}{T}(\beta_t - \beta^*), \label{eq:epsilonlbound}
\end{align}
and
\begin{align}
\E[\Delta_{t+1}] &= \Delta_t -\Delta_t (\beta_t-\beta^*) s \nonumber \\
&= (n-1)(\beta_t -\beta_t^2 s +\beta_t\beta^*s)  \nonumber\\
 &\ge (n-1)(\beta_t -\beta_t s +\beta_t\beta^*s)  \nonumber\\
  &= (n-1)\beta_t(1 - s +\beta^*s) \nonumber \\
  &\ge (n-1)\beta^*(1 - s +S/T)\ .  \label{eq:meanlbound}
\end{align}

Now, using the Chernoff bound, we have
\begin{equation*}
\Pr(\Delta_{t+1}\leq \Delta^*) = \Pr(\Delta_{t+1}\leq (1-x)\E[\Delta_{t+1}]) \ \le \exp\big(-\tfrac13 x^2 \E[\Delta_{t+1}] \big),
\end{equation*}
so when $\beta_t \ge \beta^{*} +n^{-1/2+\varepsilon}$, using \eqref{eq:epsilonlbound} and \eqref{eq:meanlbound}  we get
\begin{equation}
\label{eq:prob3}
\Pr(\Delta_{t+1}\leq \Delta^*) \ < \exp\big(-\alpha n^{2\varepsilon}\big),
\end{equation}
which is exponentially small for any constant $\varepsilon>0$, where $\alpha = \beta^*(1-S)^2(1-s+S/T)/(3T^2)$.
Stochastic domination  then implies that, when $\delta_t \ge \delta^{*} +n^{1/2+\varepsilon}$, $\delta_t$ decreases steadily,
except for some exponentially small probability.
By symmetry, we have that if $\delta_t \le \delta^{*} -n^{1/2+\varepsilon}$, $\delta_t$ increases steadily, except for exponentially small probability.

Let $\sigma_t=\delta_t/(n-1)$, and
\[ \sigma^*\ =\ \frac{(nS-1)}{(n-1)(S+T-1)}\ \sim\ \frac{S}{S+T-1},\]
so \eqref{eq:prob1} becomes
\begin{equation}
\label{eq:prob2}
 p_t\ =\  (1-\sigma_t+1/(n-1))(\sigma_t -\sigma^*)\frac{(S+T-1)}{T}\ =\ q_t(\sigma_t-\sigma^*),
\end{equation}
where \[(1/(n-1)) \frac{(S+T-1)}{T} \leq q_t\leq (1-\sigma^*+1/(n-1))\frac{(S+T-1)}{T},\] so $q_t$ is bounded away from $0$ and $1$. (Similarly, we have $\rho_t\ =\  z_t(\sigma^*-\sigma_t)$, where $0 < z_t\leq \sigma^*(S+T-1)/T < 1$.)
Also \eqref{eq:drift1} becomes
\begin{equation}
\label{eq:drift2}
\E[\sigma_{t+1}] = \sigma_t  - \sigma_t(1-\sigma_t+1/(n-1)) (\sigma_t-\sigma^*)(S+T-1)/T\,.
\end{equation}
We suppose $\sigma_0=1$ (i.e. $n-1$ defectors at $t=0$). Next we determine
the time it takes for $\sigma_t$ to drop into the range  $\sigma^{*} \pm n^{-1/2+\varepsilon}$.

First, consider $\sigma_t\geq \tfrac12(1+\sigma^*)$. Then \eqref{eq:drift2} implies
\begin{equation}
\label{eq:inequ1}
\E[\sigma_{t+1}]\ \leq\ \sigma_t  - \lambda(1-\sigma_t+1/(n-1))\ \leq\ \sigma_t  - \lambda(1-\sigma_t) ,
\end{equation}
where $\lambda={\sigma^*}^2(S+T-1)/2T$. We have immediately
$\E[\sigma_1] \leq 1-\lambda/(n-1)$. Assume by induction that
$\E[\sigma_t] \leq 1-\lambda(1+\lambda)^{t-1}/(n-1)$. Then, from \eqref{eq:inequ1},
\[ \E[\sigma_{t+1}]\ \leq\ \sigma_t  - \lambda(1-\sigma_t)\ \leq\
1-\frac{\lambda(1+\lambda)^{t-1}}{n-1}-\frac{\lambda^2(1+\lambda)^{t-1}}{n-1}
\ =\ 1-\frac{\lambda(1+\lambda)^t}{n-1},\]
continuing the induction. Thus, at time $t_0=O(\log n)$, we will have $\E[\sigma_{t_0}]< \tfrac12(1+\sigma^*)$. Then \eqref{eq:drift2} implies
\begin{equation}
\label{eq:inequ2}
\E[\sigma_{t+1}-\sigma^*]\ \leq\ \sigma_t-\sigma^*  - \mu(\sigma_t-\sigma^*)\ =\ (1-\mu)(\sigma_t-\sigma^*) ,
\end{equation}
where $\mu=(1-\sigma^*)\sigma^*(S+T-1)/2T$. We have $\E[\sigma_{t_0}-\sigma^*]\leq \tfrac12(1-\sigma^*)$, and $t_0=O(\log n)$,
so, at time $t_1=O(\log n)$, we will have $\E[\sigma_{t_1}]\leq\sigma^*+n^{-1/2+\varepsilon}$, for any constant $\varepsilon>0$. We will also have $\sigma_{t_1}>\sigma^*$, except for exponentially small probability, from \eqref{eq:prob3}.

When $t>t_1$, we require more careful analysis.
As $\sigma_{t+1}$
can be greater or less than $\sigma^{*}$, we will consider the quantity $(\sigma_{t+1}-\sigma^*)^2$.
That is,
\begin{equation*}
\E[(\sigma_{t+1}-\sigma^*)^2]\ =\ \E[(\sigma_{t+1}-\E[\sigma_{t+1}])^2]+\big(\E[\sigma_{t+1}]-\sigma^*\big)^2 .
\end{equation*}
Here, $\E[(\sigma_{t+1}-\E[\sigma_{t+1}])^2]$ can be determined from the variance of Bin$(\delta_t, p_t)$
if $\sigma_t > \sigma^*$, and from the variance of Bin$(n-\delta_t, \rho_t)$ if  $\sigma_t < \sigma^*$,
by scaling. Below we use the fact that the variance of Bin$(m,p)$ is $mp(1-p)\leq mp$.  Now,
using \eqref{eq:prob2}, \eqref{eq:inequ2} and that the drift around $\sigma^{*}$ is symmetrical, we get
\begin{align*}
\E[(\sigma_{t+1}-\sigma^*)^2]\ &=\ \E[(\sigma_{t+1}-\E[\sigma_{t+1}])^2]+\big(\E[\sigma_{t+1}]-\sigma^*\big)^2\\
&\leq\ \frac{e_t |\sigma_t-\sigma^*|}{(n-1)^2}+(1-\mu)^2(\sigma_t-\sigma^*)^2, \quad \mbox{where $e_t = \delta_t q_t$ or $(n-\delta_t)z_t$}\\
&< \ \frac{|\sigma_t-\sigma^*|}{(n-1)}+(1-\mu)^2(\sigma_t-\sigma^*)^2\\
&\leq\ (\sigma_t-\sigma^*)^2\Big(\frac{c_0}{(n-1)|\sigma_t-\sigma^*|}+(1-\mu)^2\Big)\\
&\ < (1-\mu)(\sigma_t-\sigma^*)^2,
\end{align*}
for some constant $c_0$, provided
\[|\sigma_t-\sigma^*|\ >\ \frac{c_0}{(n-1)\mu(1-\mu)}\ \sim\ c_1/n,\]
for some constant $c_1$. Observe that this applies both when $\sigma_t\geq\sigma^*$ and $\sigma_t<\sigma^*$.

Thus, at time $t_2=O(\log n)$, we will have either $|\sigma_{t_2}-\sigma^*|\leq c_1/n$ or $\E[(\sigma_{t_2}-\sigma^*)^2]\leq c_1^2/n^{k+2}$, for any constant $k$. In the latter case, using Markov's inequality, we have
\[ \Pr\big(|\sigma_{t_2}-\sigma^*|\geq c_1/n\big)\ =\ \Pr\big((\sigma_{t_2}-\sigma^*)^2\geq c_1^2/n^2\big)\ \leq\ 1/n^k.\]
So, in either case, we will have $|\sigma_{t_2}-\sigma^*|\leq c_1/n$ with probability at least $1-1/n^k$, for any $k$.

Thus, at $t=t_2$, we have $|\sigma_t-\sigma^*|=c_2/(n-1)$, for some constant $c_2\geq 1$. Then we have $|\delta_t-(n-1)\sigma^*|=(n-1)|\sigma_t-\sigma^*|= c_2$. Also
\[ p_t\ \sim\  \frac{c_2 (1-\sigma^*+1/(n-1))(S+T-1)}{(n-1)T}\ \sim\  \frac{c_3}{n-1},\]
for some constant $c_3$. So $\delta_{t}- \delta_{t+1}$ is approximately binomial Bin$\big((n-1)\sigma^*,c_3/(n-1)\big)$, which is approximately Poiss$\big(c_3\sigma^*\big)$.

Suppose $(n-1)\sigma^*$ is an integer $\ell$. Then $\Pr(\delta_{t+1}=\ell)$ is a constant. If this occurs, we have $p_t=0$, so the process will remain at $\delta_t=\ell$ forever. If it does not occur, we will have $|\delta_{t+1}-\ell|=O(\log n)$ with probability at least $1-n^{-k}$, for any $k$. Thus, after a further time $O(\log\log n)$, we will again have that $\Pr(\delta_{t+1}=\ell)$ is a constant. After $O(\log n)$ repetitions of this, we will have $\Pr(\delta_t=\ell)=1-n^{-k}$, for any $k$. The total time for this to occur will be $O(\log n\,\log\log n)$.

If $(n-1)\sigma^*$ is not an integer, then $|\sigma_t-\sigma^*|=\Omega(1/n)$ for all $t$. So $\Pr\big(\delta_{t+1}\in\{0,n\}\big)=n^{-O(n)}$, for all $t$. Thus, after time $n^{O(n)}$, we will have $\delta_{t+1}\in\{0,n\}$ with high probability.
\end{proof}

\begin{rem}
If $\delta^{*} < n/2$, the game eventually converges
to defection with higher probability, whereas, if $\delta^{*} < n/2$, the game eventually converges to cooperation with higher probability.
To see this, suppose $\delta^{*} < n/2$. Then, due to the symmetry about $\delta^{*}$, we have
\[\Pr[\delta_T = 0] = \Pr[\delta_T = 2\delta^{*}], \]
at time $T$. Moreover, even after reaching $2\delta^{*}$, it still takes exponential time for $\delta_t$  to reach $n$, the all-cooperate state. Hence, the game will reach the all-defect state with higher probability. Similarly,  the game will reach cooperation  with higher probability when  $\delta^{*} > n/2$.
\end{rem}

\section{Simulations}
\label{sec:simulation}
Since our results are largely asymptotic, we have also simulated the imitation update
rule on the cycle and the complete graph. The results are presented in this section.

\subsection{Cycle graph}
\begin{figure}[ht!]
\centering
\subfigure[When game started with $10\%$ cooperators at random positions.] {
\label{fig:CoopL100_InitialCoop10}
\includegraphics[width=55mm]{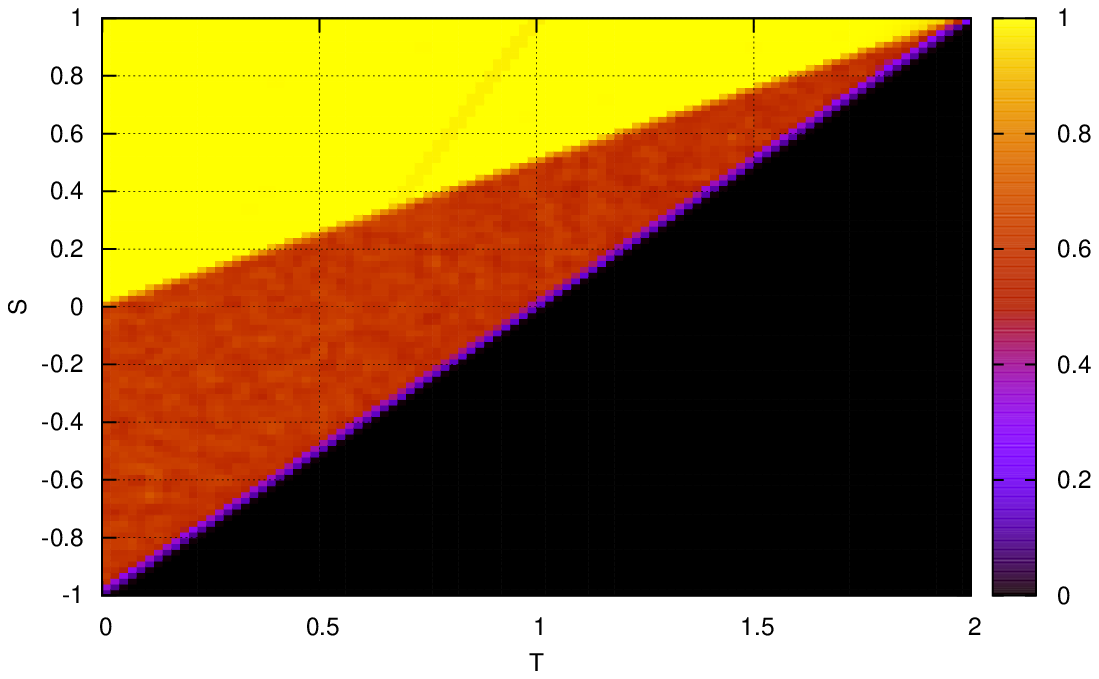}}%
\hspace{1cm}
\subfigure[When game started with $25\%$ cooperators at random positions.] {
\label{fig:CoopL100_InitialCoop25}
\includegraphics[width=55mm]{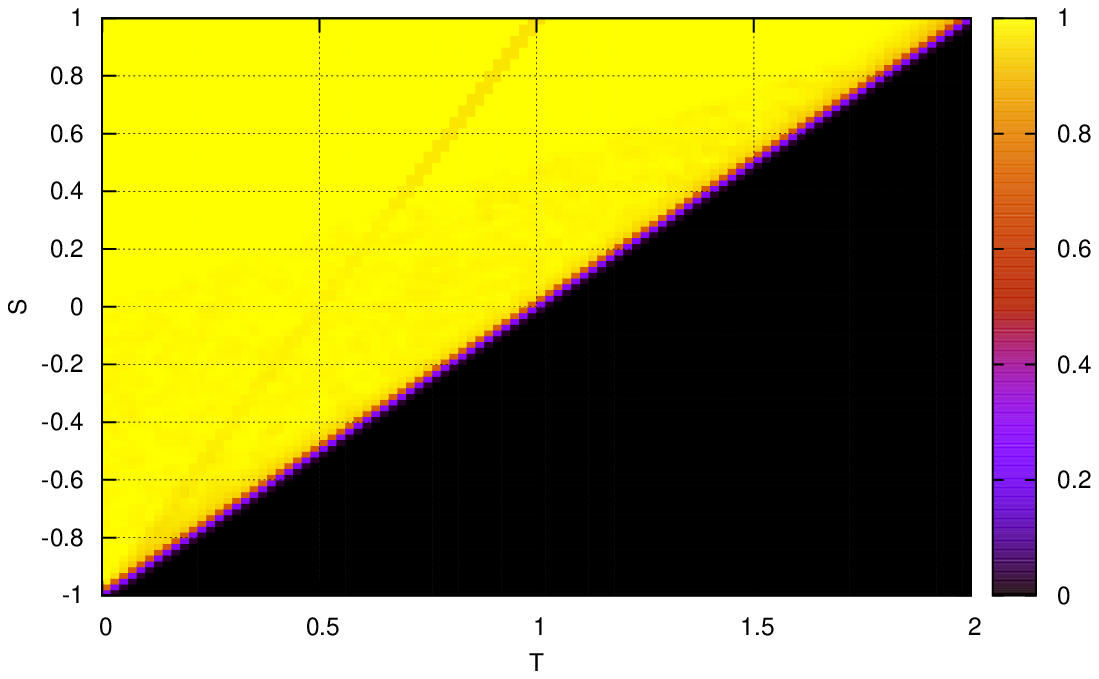}}%
\caption[Fraction of cooperators taken after 10000 generations when the game is played on a cycle of length 100.]
{Fraction of cooperators on the cycle of length 100  after 10000 generations plotted as a contour for varying $S$ and R.}
\label{fig:CoopL100_Initial_10_25} 
\end{figure}

\begin{figure}[ht!]
\centering
\subfigure[After 1000 generations.] {
\label{fig:CoopL100_G1000}
\includegraphics[width=55mm]{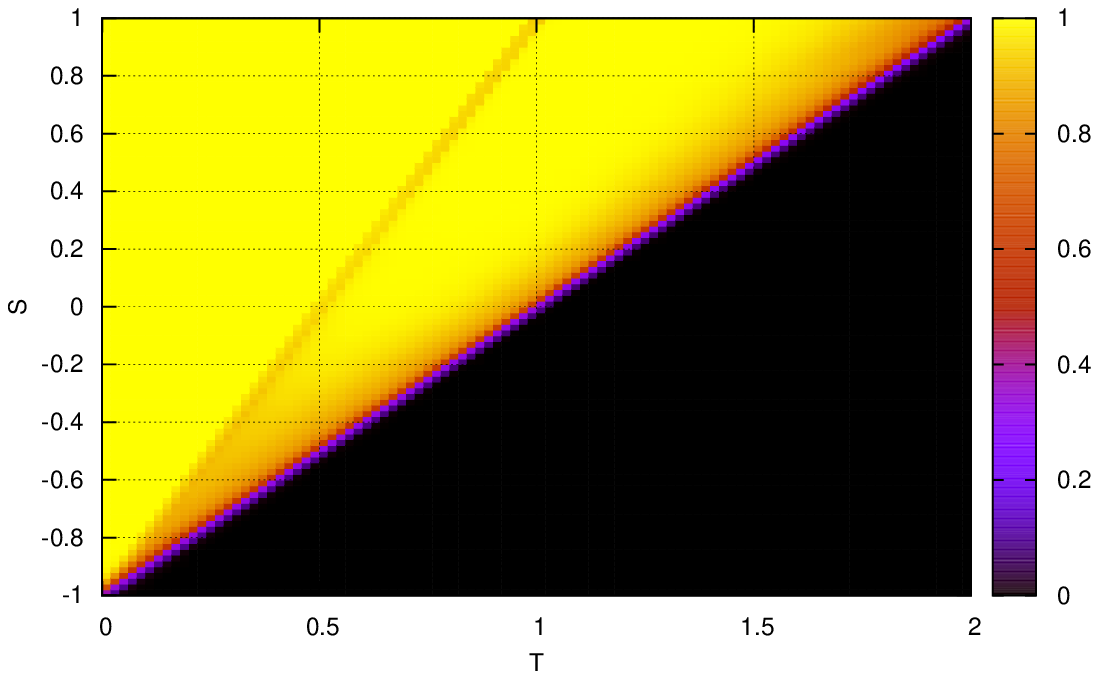}}%
\hspace{1cm}
\subfigure[After 10000 generations.] {
\label{CoopL100_G10000}
\includegraphics[width=55mm]{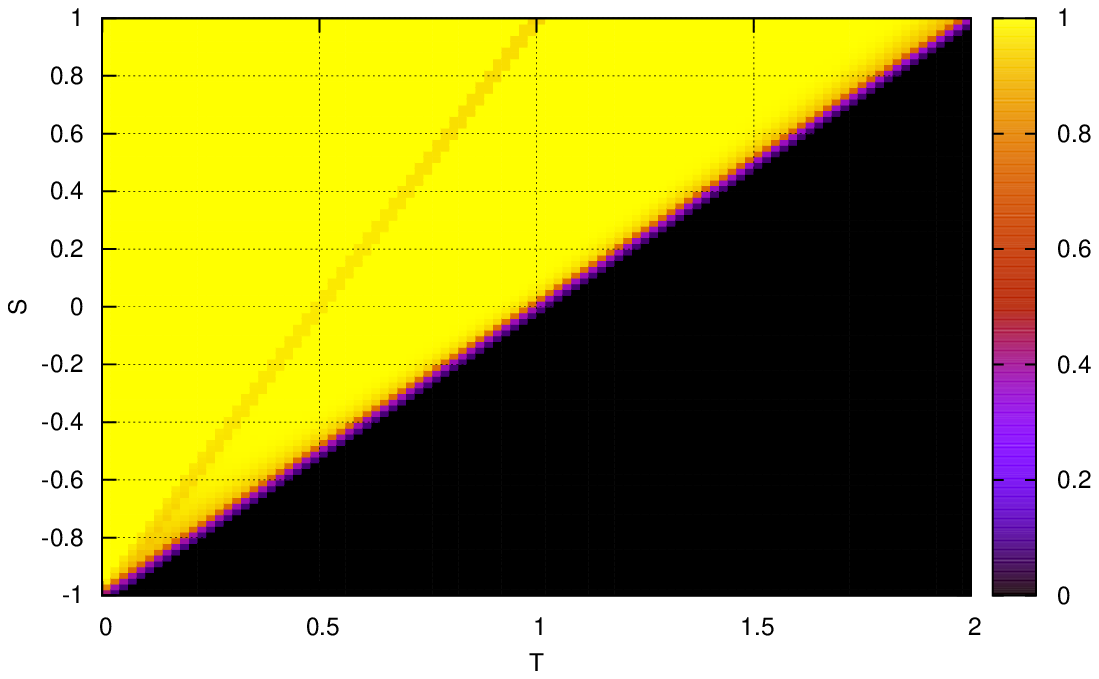}}%
\caption[Fraction of cooperators when the game is started with  $50\%$ cooperators on a cycle of length 100.]
{Fraction of cooperators on the cycle of length 100 plotted as a contour for varying $S$ and R, when the game is
started with $50\%$ cooperators at random positions.}
\label{fig:CoopL100_Initial_50} 
\end{figure}

 Figures~\ref{fig:CoopL100_Initial_10_25} and \ref{fig:CoopL100_Initial_50}
 show the results obtained for the cycle.
Each data point in these figures represents an average of 100 repetitions.
The simulations were run for $100 \times 100$ different values of $S$ and T,
with an initial configuration having cooperators uniformly distributed on the cycle.

Figures~\ref{fig:CoopL100_Initial_10_25} shows the fraction of cooperators present
on a cycle of length 100 after 10000 generations. When the game is started with
$10\%$ cooperators, the game converges to
the all-defect state if $T > S+1$ (region A); whereas, to  the all-cooperate state if
$T > (S+1)/2$ (regions C, E, and F). These results agree with our analytical results in
Section~\ref{sec:A} and Section~\ref{sec:emergenceofcoop}.
Now, in the region where $2S \le T \le S+1$ (regions B and D),
the average cooperators after 10000 generations is around $60\%$. Closer examination of the results shows that this is
because the game reaches the all-defect state fast around $40\%$ of the time and the all-cooperate state fast
around $60\%$ of the time.
This is not surprising, since when the cooperators proportion is as low
as $10\%$, the required barriers might not be present at the beginning of the game.
More precisely, recall that region B
needs 4 consecutive ($4_\textrm{c}$-run) cooperators for a guaranteed convergence to cooperation.
When $n=100$ and $p_\textrm{c}= 0.10$, the expected number of such barriers
present at the beginning of the game is very low ($0.01$). In region D, the
smallest barrier is two consecutive cooperators ($2_\textrm{c}$-run). Here again,
the expected number of barriers is low ($1.00$).

This raises an interesting question: how can the game converge to cooperation
all the time in C and E, which also need a barrier of 4 and 2 consecutive cooperators, respectively?
The answer is simple. In C and E, a singleton cooperator ($1_\textrm{c}$-run) can spread if
it has two defectors adjacent to it. However, we did not consider a singleton
cooperator as a barrier in our analysis in Section~\ref{sec:emergenceofcoop}. Because, even
in these regions, when a singleton cooperator ($1_\textrm{c}$-run) is adjacent to a singleton
defector ($1_\textrm{d}$-run), it might become a defector.

With this explanation, it is not surprising that, when the game is started
with $25\%$ cooperators, the convergence towards the all-cooperate state  is more frequent
(more than $98\%$) when $2S \le T \le S+1$.  This is shown in Figure~\ref{fig:CoopL100_InitialCoop25}.
This is even more apparent in Figure~\ref{fig:CoopL100_Initial_50} which shows the contour plotted for the game started
with $50\%$ cooperators. In this case, there are clearly only two different behaviours,
except for some minor border effects. It can be observed that the average cooperators after
1000 generations is little less than $100\%$ in some places. But, it is clear
that the game converges to cooperation before 10000 generations even in these cases.

Finally, all the plots show some minor effects along the line $T = \tfrac{S+1}{2}$ and some
 noticeable effect along the line $T=S+1$. This is due to the differences in switching probabilities
 along these borders. Again, this
concurs with our  analytical results in Section~\ref{sec:Borders}, that cooperators and defectors
can coexist in these cases.

\subsection{Complete graph}

\begin{figure}[ht!]
\centering
\subfigure[Initial cooperators = $25\%$.] {
\label{fig:K_100_InitialCoop25}
\includegraphics[width=50mm]{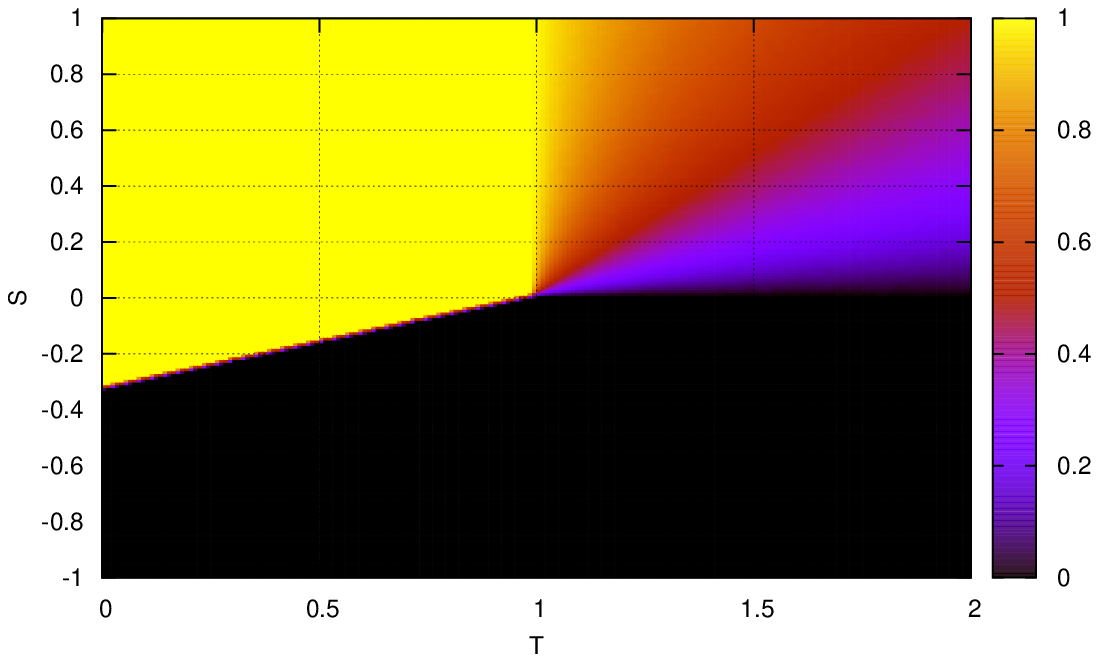}}%
\subfigure[Initial cooperators = $50\%$.] {
\label{fig:K_100_InitialCoop50}
\includegraphics[width=50mm]{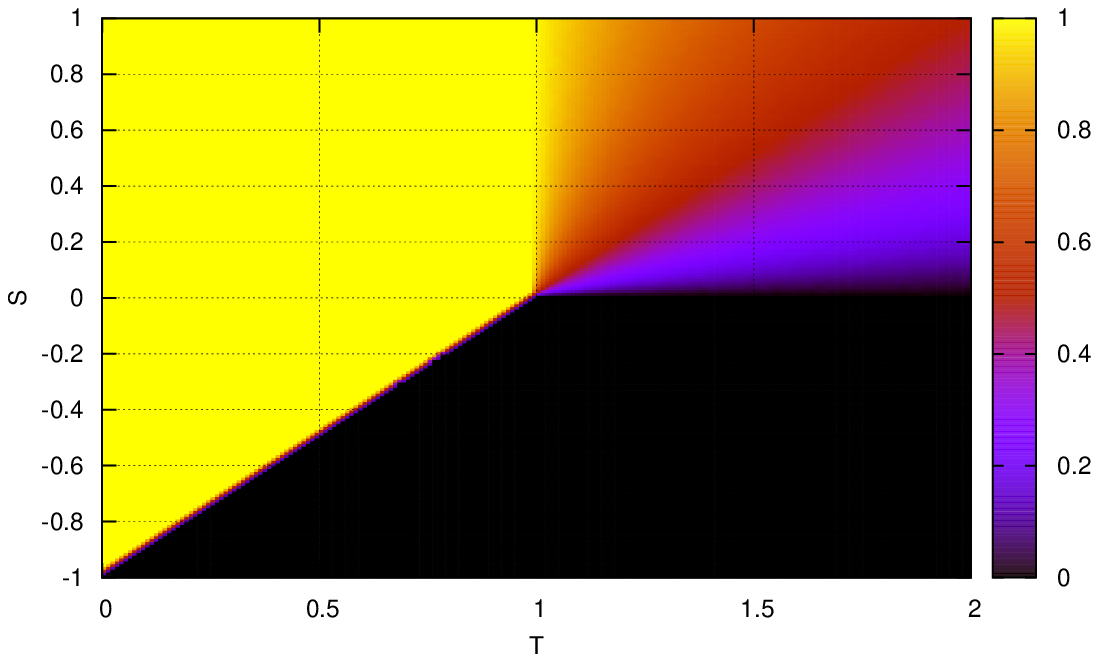}}%
\subfigure[Initial cooperators = $75\%$.] {
\label{fig:K_100_InitialCoop75}
\includegraphics[width=50mm]{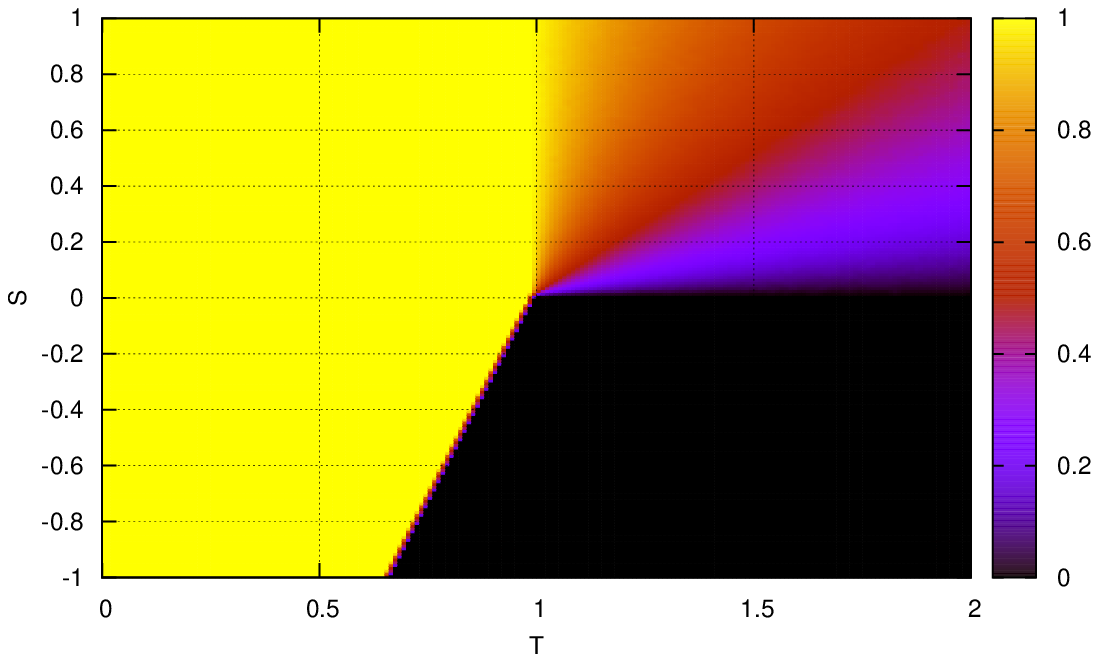}}%
\caption[Fraction of cooperators after 10000 generations when the game is played on a complete graph of size 100.]
{Fractions of cooperators after 10000 generations  on the complete graph of size 100 with different initial
configurations plotted as a contour for varying $S$ and R.}
\label{fig::K_100_Simulation} 
\end{figure}

For the complete graph, as shown in Section~\ref{sec:completeGraph}, the
exact number of cooperators present at the beginning of the game
determines whether the game converges to defection or
cooperation in the quadrant SH. For this reason, each simulation
was started with a fixed number of players. The result obtained
are plotted as a contour for varying $S$ and $T$ in Figure~\ref{fig::K_100_Simulation}.

Here again, the results agree with our rigorous analysis.
In SG region, the game did not reach a steady state even after 10000 steps.
Investigation of the data reveals that the percentage of the cooperator
after 10000 steps is close to the value of $\theta^{*}$ obtained in the analysis
(e.g.\ $\theta^{*} = 0.49$ when $n=100, T= 1.5,$ and $S=0.5$). Also, it is clear
that the initial cooperator percentage determines the line in SH region which
divides the region with the all-cooperate steady state from the region with
the all-defect steady state (e.g.\ the line is $T \approx S+ 1$ when $n=100$ and $\theta_0=0.5$).

Finally, we note that Santos \emph{et al.} \cite{Santos2006a} also produced simulations
for complete graphs. Our simulations agree with theirs.

\section{Discussion}
\label{sec:discussion}
We have studied the imitation update rule in two extreme cases of
graph topology. First,  for the cycle, we have proved that all games
converge to either cooperation or defection fast. More precisely,
if $T > S+1$, the games converge to the all-defect state fast; and if
$T < S+1$, the games converge to the all-cooperate state fast. Even within
these regions, the convergence rate for the games is different
for different values of the payoffs. It is notable from the analytical
results that the closer the point $(S,T)$ is to the line $T= S+1$, the
slower the convergence. The fact that the cooperators
cannot form a barrier (or a \emph{cluster} as Nowak and May~\cite{Nowak2006} call it) in PD region seems to suggest
that one dimensional graphs cannot help the evolution of cooperation through network reciprocity.

We highlight the fact that, for the complete graph also,
all four games converge to cooperation
or defection. But the rate of convergence is
exponentially slow in $n$ for the SG game.
We note that experimental
studies, including~\cite{Nowak2006,Santos2006a}, have wrongly concluded
that cooperators and defectors can coexist indefinitely.
Our results show that this is true for the complete graph only for some very particular values of $S$ and $T$.
In the SG region, as we have proved, the number of cooperators oscillates
around a value, $\delta^*$, for an exponential amount of time. We
showed that convergence to cooperation is more likely after an exponential
time if $\delta^*$ is closer to $n$, and the convergence to defection
 is more likely if $\delta^*$ is closer to $0$. For some special values of
$S$ and $T$, where $\delta^*$ is an integer,
$\delta^*$ cooperators is a steady state. In this state,
cooperators and defectors earn equal payoff. We have proved that, for these special
values of $S$ and $T$, the convergence to this steady state happens fast.

\section{Conclusions and open problems}
\label{s:conclusion}
We have shown that indefinite coexistence of cooperators and defectors is
impossible on the cycle, except for some special values of $S$ and $T$.
More precisely, the coexistence is only possible
for games where $T= S+1$ or $T=(S+1)/2$. Furthermore, we have
shown that, for all the games studied, a steady state is reached for cycles in polynomial
time. That is, cooperation emerges rapidly when $S + 1 > T$ and
$T \ne (S+1)/2$; defection emerges rapidly when $S + 1 < T$; and, a steady state with cooperators and defectors
is reached rapidly  when $T = S + 1 $ and  $T = (S+1)/2$. We also analysed the imitation strategy
on complete graphs. The analysis reveals that defection emerges fast for Prisoner's
Dilemma game, and cooperation emerges fast for Harmony game. In the Stag Hunt game, either cooperation or
defection emerges fast depending on the initial proportion of cooperators.
In the Snowdrift game, a metastable state is reached fast. In this state, the proportion of
cooperators fluctuates around a fixed value for exponential time, before converging
to cooperation or defection.

It remains as an open question whether there are graphs other than the cycle
on which cooperators and defectors cannot coexist. An interesting extension
of this work would be to study rigorously the imitation strategy on other
graphs, such as trees and grids. In particular, based on simulations presented in~\cite{Roca2009b}, regular lattices seem to show very similar, if not the same, behaviour to the complete graphs for the whole $ST$ region.
But it does not appear that a similar analysis can be used to prove this.

\bibliographystyle{amsplain}
\bibliography{Bibs}

\end{document}